\documentclass[a4paper,11pt]{article}
\pdfoutput=1 
\usepackage{jheppub}
\usepackage{graphicx}
\makeatletter

\renewcommand*{\p@subsection}{}

\renewcommand*{\p@subsubsection}{}
\makeatother

\usepackage{amsmath}
\usepackage{pstricks}
\usepackage{color}
\usepackage{xcolor}
\usepackage{graphicx}
\usepackage{dcolumn}
\usepackage{bm}
\usepackage{slashed}
\usepackage{subcaption}
\usepackage{float}
\usepackage{hyperref}
\usepackage{caption}
\usepackage{threeparttable}
\usepackage{jheppub}

\newcommand{\bea}{\begin{eqnarray}}
\newcommand{\eea}{\end{eqnarray}}
\newcommand{\beq}{\begin{eqnarray}}
\newcommand{\eeq}{\end{eqnarray}}

\def\bit{\begin{itemize}
  }
  \def\eit{\end{itemize}
  }

\begin{document}

\preprint{}

\title{\centering WIMP Cogenesis for Asymmetric Dark Matter \newline and the Baryon Asymmetry}

\author[a]{Yanou Cui,} 
 \emailAdd{yanou.cui@ucr.edu}
\author[a]{Michael Shamma}
 \emailAdd{michael.shamma@email.ucr.edu}
\affiliation[a]{Department of Physics and Astronomy, University of California, Riverside, CA 92521, USA}

\date{\today}

\abstract{
We propose a new mechanism where asymmetric dark matter (ADM) and the baryon asymmetry are both generated in the same decay chain of a metastable weakly interacting massive particle (WIMP) after its thermal freezeout. Dark matter and baryons are connected by a generalized baryon number that is conserved, while the DM asymmetry and baryon asymmetry compensate each other. This unified framework addresses the DM-baryon coincidence while inheriting the merit of the conventional WIMP miracle in predicting relic abundances of matter. Examples of renormalizable models realizing this scenario are presented. These models generically predict ADM with sub-GeV to GeV-scale mass that interacts with Standard Model quarks or leptons, thus rendering potential signatures at direct detection experiments sensitive to low mass DM. Other interesting phenomenological predictions are also discussed, including: LHC signatures of new intermediate particles with color or electroweak charge and DM induced nucleon decay; the long-lived WIMP may be within reach of future high energy collider experiments.}

\maketitle
\thispagestyle{empty}
\newpage


\section{Introduction}\label{sec:intro}
\setcounter{page}{1}

The cosmic origins of baryon and dark matter (DM) abundances have been long-standing puzzles in particle physics and cosmology. In most proposals, the explanation for DM and baryon abundances today are treated with separate mechanisms. Meanwhile, the observation that their abundances are strikingly similar, $\Omega_{DM}/\Omega_{B}\approx5$ \cite{planck}, presents a coincidence problem, and suggests a potential connection between DM and baryons in the early Universe. These together form a \textit{triple puzzle} about matter abundance in our Universe.

The WIMP miracle, i.e. through thermal freezeout, DM with weak-scale interactions and masses gives the correct DM abundance today, has been a leading paradigm for DM model-building. The WIMP paradigm does not address the DM-baryon coincidence. Meanwhile, conventional WIMPs have been increasingly constrained by indirect/direct detection and collider experiments \cite{darkside50, indirect, Mitsou:2014wta}. This has led to the proliferation of exploring alternative DM candidates beyond of the WIMP paradigm. Asymmetric dark matter (ADM) \cite{Nussinov:1985xr, Barr:1990ca, Kaplan:1991ah, Kaplan:2009ag, Zurek:2013wia, adm} is one alternative to WIMP DM, inspired by the DM-baryon ``coincidence". In this framework, the DM particle is distinct from its antiparticle, and an asymmetry in the particle-antiparticle number densities is generated in the early universe. Subsequently, the symmetric component is annihilated away by efficient CP-conserving interactions, leaving the asymmetric component to dominate the DM density today. The core idea of ADM is based on relating DM and baryons/leptons, through shared interactions in the early Universe. The generation of the initial DM or baryon asymmetry for ADM often requires a separate baryogenesis-type of mechanism. In general ADM models do not possess the attractive merit of the WIMP miracle in predicting the absolute amount of matter abundance.

WIMP DM and ADM are both appealing proposals that address some aspect of the aforementioned triple puzzle about matter. However, it is intriguing to explore the possibility of a unified mechanism that combines their merits and addresses all three aspects of the puzzle simultaneously. Recently a few attempts have been made in this direction \cite{wimpyBG, McDonald:2011zza, Davidson:2012fn, wimpyBG2, Cui:2015eba, Farina:2016ndq, Racker:2014uga, Cui:2013bta}. Among these existing proposals, \cite{McDonald:2011zza} is highly sensitive to various initial conditions, while both \cite{Davidson:2012fn} and WIMP DM annihilation triggered ``WIMPy baryogensis" \cite{wimpyBG} have sensitivity to washout details. The mechanism of ``Baryogenesis from Metastable WIMPs" \cite{wimpyBG2} was then proposed as a alternative where the prediction is robust against model details: the baryon asymmetry is generated by a long-lived WIMP that undergoes CP- and B-violating decays after the thermal freezeout of the WIMP. Such models also provide a strong cosmological motivation for long-lived particle searches at the collider experiments and have become a benchmark for related studies \cite{Cui:2014twa, Cui:2016rqt, ATLAS:2019ems}. However, the original model of Baryogenesis from Metastable WIMPs does not involve specifics of DM, only assuming that DM is another species of WIMP that is stable, and thus the DM-baryon coincidence is addressed by a generalized WIMP miracle which is not fully quantitative. From model building perspective it would be more desirable to further develop a framework which incorporates the merits of \cite{wimpyBG2} as well as the details of DM, and predicts a tighter, more precise connection between $\Omega_{\rm DM}$ and $\Omega_B$. There are two possible directions to pursue for this purpose: consider a WIMP DM that closely relates to the metastable baryon-parent WIMP in \cite{wimpyBG2} (e.g. in the same multiplet or group representation), or consider a further deviation from \cite{wimpyBG2} where the post-freezeout decay of a grandparent WIMP generates both DM and baryon asymmetries, thus DM falls into the category of ADM. In this work we explore the latter possibility, which we naturally refer to as ``WIMP cogenesis". The WIMP of our interest is of conventional weak scale mass or moderately higher (up to $\sim10$ TeV). We aim at constructing a viable WIMP cogenesis model with the following guidelines: 
\begin{itemize}
\item UV complete, only involves renormalizable interactions; 
\item ADM $X$ and baryon asymmetries are generated in the same decay chain (instead of two different decay channels with potentially arbitrary branching ratios) so as to have the least ambiguity in predicting their ``coincidence";
\item The model possesses a generalized baryon/lepton number symmetry $U(1)_{B (L)+kX}$ that is conserved. 
\end{itemize}
$k$ is a model-dependent $O(1)$ rational number that parametrizes the ratio of ADM number to baryon (lepton) number produced in the decay chain. These first two guidelines distinguish our model from some other existing ADM proposals based on massive particle decay, such as \cite{dmfrombaryonasymmetry, admfromlepto, hylogenesis}. In particular, the second guideline leads to a neat prediction of the ADM mass: 
\begin{equation}\label{eq:admmass}
m_{X}=c_{s}\frac{1}{k}\frac{\Omega_{X}}{\Omega_{B}}m_{n},
\end{equation}
where $m_n\approx1$ GeV is the neutron mass, $k=2$ in the benchmark models we will demonstrate, the baryon distribution factor $c_{s}=\frac{n_{B}}{n_{B-L}}\sim O(1)$ depends on whether the EW sphaleron is active when the decays occur, and will be elaborated in Sec.~\ref{sec:model}. Given that $\frac{\Omega_{X}}{\Omega_{B}}\approx5$ from observation, Eq.~\ref{eq:admmass} generally predicts $m_X$ in the GeV range. This possibility of producing DM and baryons in the same decay chain was suggested in the warped unification scenario \cite{Agashe:2004bm}, while concrete examples remain to be seen.
The third guideline, i.e., the idea of DM and baryon sharing a conserved global baryon number symmetry is also seen in e.g., \cite{Cui:2011qe, hylogenesis, Fonseca:2015rwa, Elor:2018twp}. 

\begin{figure} 
\includegraphics[width=\textwidth,trim=50 630 50 50, clip]{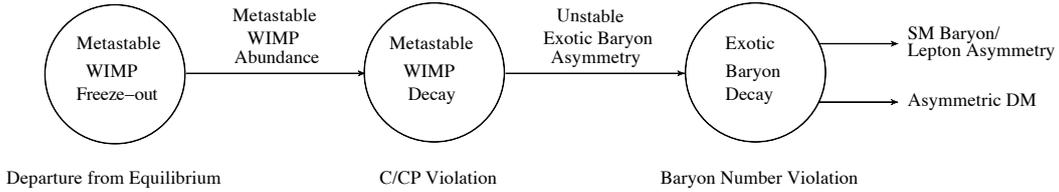}
\caption{Schematic diagram outlining the key stages in WIMP cogenesis mechanism. Each dynamical stage of WIMP cogenesis, shown in the bubbles, satisfies one of the Sakharov conditions. }
\label{fig:cartoon}
\end{figure}

The schematic idea of this new mechanism is illustrated in Fig.~\ref{fig:cartoon}, which consists of a sequence of three stages that satisfy each of the three Sakharov conditions in order.
\begin{enumerate}
\item Metastable WIMP freezeout. The out-of-equilibrium condition is automatically satisfied as a consequence of the WIMP freezeout. This step establishes a ``would-be" WIMP miracle relic abundance predicted for the grandparent WIMP that will be inherited by $\Omega_{X}$ and $\Omega_B$ when the WIMP decays: 
\begin{eqnarray}\label{eq:genres}
\Omega_{B}\sim\Omega_{X}&\approx&\epsilon_{CP}\frac{m_{B(X)}}{m_{\rm WIMP}}\Omega_{\rm WIMP}^{\tau\rightarrow\infty}\\\nonumber
&\approx&0.1\epsilon_{CP}\frac{m_{B(X)}}{m_{\rm WIMP}}\frac{\alpha^2_{\rm weak}/(\rm TeV)^2}{\langle\sigma_{\rm ann, WIMP} v\rangle}.
\end{eqnarray}
\item C- and CP-violating decay of the WIMP to intermediate states of exotic baryons/leptons. This occurs well after the freezeout and before BBN. The asymmetry between $B$ and $\bar{B}$, or between DM and anti-DM originates from this stage.
\item The decay of the intermediate exotic baryons/leptons into SM baryons/leptons and ADM. While this stage conserves the generalized $U(1)_{B (L)+X}$, the SM B-number symmetry is violated.
\end{enumerate}

The rest of the paper is organized as follows. In Section \ref{sec:quark}, we consider a model where the WIMP decay products are SM quarks and ADM leading to direct baryogenesis, where the related general formulations and numerical results will be given. Section \ref{sec:lepton} introduces a leptogenesis model where the WIMP directly decays to leptons and ADM, which induces the baryon asymmetry by sphaleron effect provided that the decay occurs before EW phase transtion. Experimental signatures and constraints are discussed in Section \ref{sec:pheno}. Section \ref{sec:fin} concludes this work.  

\section{WIMP Decay to Baryons and ADM} 
\label{sec:quark}
In this section, we explore a specific model which directly produces a baryon asymmetry along with ADM via SM B-violating interactions. The fields and interactions are introduced followed by discussions on how Sakharov conditions are met by their interactions and the related cosmological evolution. This section ends with numerical analyses of the parameter space for these types of models. 
\begin{figure}[h]
\centering
\includegraphics[width=0.4\textwidth,trim=0 150 50 150, clip]{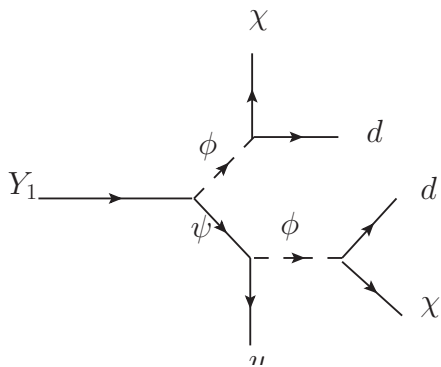}
\caption{Feynman diagram of the WIMP decay chain producing baryon and DM asymmetries.}
\label{fig:quarksteps}
\end{figure}

\subsection{Model Setup}\label{sec:model}
To implement the picture discussed in the introduction, we extend the SM with the following Lagrangian: 
\begin{eqnarray}\label{eq:Lyuk}
\mathcal{L}&=&i\frac{1}{2}\bar{Y}_{1,2}\slashed{\partial}Y_{1,2}+\bar{\psi}_{i}(i\slashed{\partial}-m_{\psi})\psi_{i}+\bar{\chi}(i\slashed{\partial}-m_{\chi})\chi+(\partial^{\mu}\phi_{i})^{\dagger}(\partial_{\mu}\phi_{i})\\ \nonumber&-&m_{\phi}^{2}\phi^{\dagger}\phi -\eta_{1,2}\phi_{i}\bar{Y}_{1,2}P_{R}\psi_{i}-\alpha_{ii}\phi_{i}\bar{d}_{i}P_{L}\chi^{c}-\beta_{ijk}\phi_{i}\bar{\psi}_{j}P_{R}u_{k}+\text{h.c.}
\end{eqnarray}
where $u^{i}$ and $d^{i}$ are the SM quark fields. With the chiral projectors, only right-handed quarks are relevant. The SM singlet $\chi$ is the ADM, all Yukawa couplings are generic complex numbers, and $\beta_{ijk}$ is anti-symmetric in its indices. Two Majorana fermions $Y_{1,2}$ are introduced: $Y_1$ plays the role of the WIMP grandparent for the ADM and baryon asymmetry, while $Y_2$ is essential for the interference process that enables C- and CP-violation (see Sec. \ref{sec:cpv}). Three generations of diquark scalars $\phi_{i}$ and vector-like Dirac fermions $\psi_{i}$ are the exotic baryons that are the intermediate decay products of metastable $Y_{1}$ as described in Stage-2 in Sec.\ref{sec:intro}. This Lagrangian possesses a $U(3)$ flavor symmetry under which $\psi_i, \phi_i$ transform as fundamentals. The model is thus consistent with minimal flavor violation and forbids new sources of flavor-changing neutral currents (FCNC). Note that the $U(3)$ flavor symmetry is optional for the purpose of suppressing FCNC: with couplings $10^{-7}\lesssim\alpha\lesssim0.1$, there is no effect on the prediction for matter abundances in our model, while the FCNC constraint can be satisfied. Nevertheless with $\alpha\lesssim0.1$ the potential DM direct detection signal (Sec. \ref{sec:dd}) would be too small to be observed. CP-violating $Y_{1}$ decays produce asymmetries in intermediate states $\phi$ and $\psi$ and their conjugates. These states subsequently decay to produce asymmetries between $udd$ and $\chi$ and their conjugates. For simplicity, we have taken the different flavors of $\psi$ and $\phi$ to be degenerate in mass. This will be the case throughout the rest of the paper. The Feynman diagram for the decay chain is shown in Fig. \ref{fig:quarksteps}.
\begin{figure}[b]
\centering
\includegraphics[width=0.9\textwidth, trim=0 195 0 200, clip]{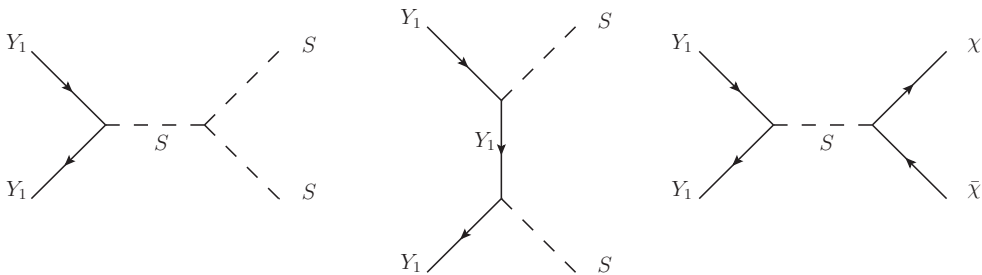}
\caption{Annihilation processes that potentially contribute to $Y_{1}$ freezeout.}
\label{fig:ann}
\end{figure}
The above symmetries allow additional interactions between the Majorana singlet and the SM through $\bar{L}HY_{1}$ which permit decays $Y_1\rightarrow Hl$. It is technically natural for this coupling to remain small such that $Y_1$ decays to $\phi\psi$ are dominant. Alternatively, the Yukawa interaction $\bar{L}HY_{1}$ is forbidden by imposing an exact $Z_{4}$ symmetry with the following charge assignments: $Y_1$ charge $-1$, $\psi,~\phi,~\chi$ charge $i$, and all SM charges are $+1$. This $Z_{4}$ symmetry also ensures the stability of asymmetric dark matter candidate $\chi$. 

\begin{figure}[]
\centering
\includegraphics[width=0.75\textwidth, trim=0 250 0 250, clip]{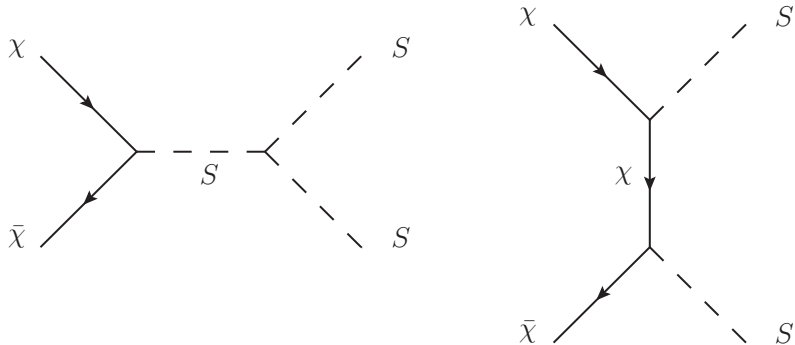}
\caption{Annihilation process that depletes the symmetric component of $\chi\bar{\chi}$.}
\label{fig:dmann}
\end{figure}
Just like with WIMP DM freezeout, there are various possibilities of $Y_1$ annihilations that can lead to a metastable WIMP abundance through thermal freezeout. Our core mechanism and result (e.g. Eq. \ref{eq:genres}) are insensitive to the detailed realization of such annihilations/freezeout. To give a concrete example, we choose to consider the simple case where $Y_1, \chi$ annihilate into SM singlet scalar S. Feynman diagrams for $Y_1$ and $\chi$ annihilation are shown in \ref{fig:ann} and \ref{fig:dmann}, respectively. Nevertheless, we are not committed to this choice: as said there are other potentially more complex possibilities to realize $Y_1$ freezeout, e.g., $Y_1$ annihilating to $\psi,~\phi,~\chi$ mediated the gauge boson of a spontaneously broken $U(1)'$  \cite{Kahlhoefer:2015bea, Cui:2017juz}. 

Specifically, for $Y_1$ freeze-out and $\chi$ symmetric component depletion, we introduce additional interactions as follows:
\begin{eqnarray}\label{eq:Ldark}
\mathcal{L}_{{\text{f.o.}}}&=&-\rho_{1,2} S\bar{Y}_{1,2}Y_{1,2}-\delta S\bar{\chi}\chi-\mu S^3
\end{eqnarray}
There may be additional interactions for $S$ such as $S\bar{\psi}\psi$ and $|\phi|^2 S^2$. These would permit further contributions to the $Y_1$ annihilation cross section. That said, the interactions in Eq. \ref{eq:Ldark} constitute a minimal, renormalizable model for $Y_1$ freeze-out and ADM symmetric component depletion. 

In order to efficiently deplete the symmetric component of $\chi$, its annihilation cross section must be $\sigma_0(\chi\bar{\chi}\rightarrow SS)\gtrsim\text{few}\times\sigma_{0,\text{WIMP}}$ \cite{Graesser:2011wi}. The cross sections, $\sigma_0$ are defined by $\langle\sigma|\vec{v}|\rangle=\sigma_0(T/m)^n$ ($n=0$ for s-wave annihilation, $n=1$ for p-wave) and $\sigma_{0,\text{WIMP}}\approx5\times10^{-26}~\text{cm}^3/\text{s}$. Following Eq. \ref{eq:Ldark}, ADM candidate $\chi$ undergoes t- and s-channel annihilations to $S$. In the $m_{\chi}\gg m_{S}$ limit, the cross section is $\sigma_0(\chi\bar{\chi}\rightarrow SS)=3\delta^2\big[24\delta^2-(20\mu\delta/3m_\chi)+(\mu^2/2m_\chi^2)]/64\pi m_{\chi}^2$. To give an example, for $\delta=0.2$, $\mu=5~\text{GeV}$ and $m_\chi=2.5~\text{GeV}$, the cross section is $\sigma_0(\chi\bar{\chi}\rightarrow SS)\approx3\times10^{-22}~\text{cm}^{3}/s$ which is sufficiently efficient to deplete the symmetric component of $\chi$. Without a symmetry protection, $S$ may decay to the SM states, e.g. through a mixing with the SM Higgs enabled by an $S|H|^2$ term.
Thus the $\chi$ asymmetry remains the dominant contribution to the DM abundance.

With these interactions we can also define a generalized global baryon symmetry $U(1)_{B+2X}$ with conserved number $G$. We will further explain the $G$ charge assignments in Sec.~\ref{sec:bv}. The generalized baryon and other charges are given in Table \ref{tab:table1}.

\begin{table}[h!]
\centering
\begin{tabular}{|l|l|l|l|l|l|l|}
\hline
        & $SU(3)_C$      & $SU(2)_L$ & $U(1)_Y$  & $U(1)_{B+2X}$ & $Z_{4}$ \\ \hline
$Y_{1,2}$ & $\bf{1}$       & $\bf{1}$  & $~~~0$         & $~~~0$              & $-1$              \\ \hline
$\psi$  & $\bf{\bar{3}}$ & $\bf{1}$  & $+2/3$         & $+1/6$            & $+i$              \\ \hline
$\phi$  & $\bf{3}$       & $\bf{1}$  & $-2/3$       & $-1/6$           & $+i$              \\ \hline
$\chi$  & $\bf{1}$       & $\bf{1}$  & $~~~0$          & $-1/2$           & $+i$              \\ \hline
$S$     & $\bf{1}$& $\bf{1}$  & $~~~0$      & $~~~0$            & $+1$              \\ \hline
$u$     & $\bf{3}$       & $\bf{1}$  & $+4/3$       & $+1/3$            & $+1$              \\ \hline
$d$     & $\bf{3}$       & $\bf{1}$  & $-2/3$        & $+1/3$            & $+1$              \\ \hline
\end{tabular}
\caption{Quantum numbers of the relevant particles in WIMP cogenesis with baryons.}
\label{tab:table1}
\end{table}

After the decay processes have taken place, efficient matter-antimatter annihilations deplete the $\bar{\chi}$ number density to near triviality. This leaves an abundance of two $\chi$'s for every unit of baryon number ($udd$). The shared interactions fix the relationship between the asymmetries of baryons and $\chi$. This then fixes the ADM $\chi$ mass according to Eq. \ref{eq:admmass}. It is apparent that $n_{B-L}/n_{DM}=1/2$ for this model. $c_{s}\equiv\frac{n_B}{n_{B-L}}$ characterizes the potential effect of redistribution among $B$ and $L$ numbers due to sphaleron interactions. If the asymmetry is produced after the electroweak phase transition (EWPT), $c_{s}=1$. If the asymmetry is produced before EWPT \cite{tasi}, SM charged particles and $\phi$, $\psi$, $\chi$ are in chemical equilibrium and their chemical potentials are related by the active gauge and Yukawa interactions as well as sphaleron processes. With the SM alone, $B-L$ is preserved, while in this model the linear combination $B-L+2X$ is conserved. Putting all these together we can solve for $c_s$. As explained in Appendix~\ref{sec:cs}, $c_s$ has a dependence on the masses of $\psi, \phi$ relative to the temperature at EWPT, $T_{\text{EWPT}}$. Given the large uncertainty in determining $T_{\text{EWPT}}$, we consider two limits of interest which would define the range of the $c_s$ values: $m_{\phi,\psi}\ll T_{\text{EWPT}}$ and $m_{\phi,\psi}\gg T_{\text{EWPT}}$. The solutions for the two limits are (details given in Appendix~\ref{sec:cs1}):
\begin{eqnarray}\label{eq:cs}
c_{s}=\frac{n_B}{n_{B-L}}=\begin{cases}
\frac{4(N_{f}+N_{H})}{14N_{f}+13N_{H}} &m_{\phi,\psi}\ll T_{\text{EWPT}}\\
\frac{8N_{f}+4N_{H}}{22N_{f}+13N_{H}} &m_{\phi,\psi}\gg~T_{\text{EWPT}}
\end{cases}
\end{eqnarray}
where $N_{f}$ and $N_{H}$ are the number of generations of fermions and number of Higgs, respectively. For matter asymmetries produced before EWPT with $N_{f}=3$ and $N_{H}=1$ Eq. \ref{eq:cs} gives $c_{s}=16/55$ for $m_{\phi,\psi}\ll T_{\text{EWPT}}$ or $c_{s}=28/79$ for $m_{\phi,\psi}\gg T_{\text{EWPT}}$. Combining these and Eq. \ref{eq:admmass}, we find that $m_{\chi}=2.5~\text{GeV}$ if the asymmetry is produced after EWPT and $m_{\chi}\approx0.72~\text{GeV}-0.89~\text{GeV}$ if produced before EWPT. 
 
Next we demonstrate how WIMP cogenesis satisfies the Sakharov conditions \cite{Sakharov} for generating a primordial asymmetry in both baryon and DM sectors. 

\subsection{WIMP Freezeout and the Generalized WIMP Miracle}\label{sec:freezeout}
The thermal freezeout of $Y_1$ provides the out-of-equilibrium condition for asymmetry generation upon the subsequent decays.

The freezeout of $Y_{1}$ proceeds through $S$ mediated annihilation to the hidden sector states $S,~\chi$. The annihilation rate is given by $\Gamma(Y_{1}Y_{1}\rightarrow\chi\bar{\chi}, SS)=n_{Y}\langle\sigma(Y_{1}Y_{1}\rightarrow\chi\bar{\chi}, SS)|\vec{v}|\rangle$. Since $Y_1$ and $\chi$ couple to scalar $S$, there is no s-wave contribution to the cross section because of helicity suppression in the fermionic case and the imposition of CP-invariance in the scalar case \cite{Bell:2017irk}. Although pseudoscalar coupling to $S$ would lift this suppression, velocity suppressed annihilation is sufficient, as we will see in Sec. \ref{sec:quarkresults}. The p-wave contributions to the $Y_{1}$ annihilation cross section are to $\chi,~S$ final states, as shown in Fig. \ref{fig:ann}.  The freezeout occurs at $T_{f.o.}$ when the $Y_{1}$ annihilation rate falls below the Hubble expansion rate, which can be estimated as follows:
\begin{eqnarray}\label{eq:fotemp}
x_{f.o}\equiv\frac{m_1}{T_{f.o.}}\simeq \ln\Bigg\{\frac{0.152g_{*}^{-1/2}M_{\text{Pl}}m_{1}\sigma_{0}}{\ln^{3/2}\big( 0.152g_{*}^{-1/2}M_{\text{Pl}} m_{1}\sigma_{0}\big)}\Bigg\}
\end{eqnarray}
where we parametrize the p-wave contributions to the thermally averaged cross-section in the limit $m_1\gg m_S$ as $\langle\sigma_{Y_{1}\text{ann}}|\vec{v}|\rangle\simeq\sigma_{0}x^{-1}$ with
\begin{eqnarray}\label{eq:sigma0}
   \sigma_{0}=\frac{3(24\rho_{1}^4-\frac{20\rho_1^3\mu}{3m_1}+\frac{\mu^2\rho_1^2}{2m_1^2}+\delta^4)}{64\pi m_{1}^2}
\end{eqnarray}
where $M_{\text{Pl}}=1.2\times10^{19}~\text{GeV}$ is the Planck mass and $g_{*}$ is the effective degrees of freedom \cite{wellsannihilation}. Since efficient depletion of the symmetric component of ADM requires $m_S\lesssim m_{\chi}\sim\mathcal{O}(\text{GeV})$ while $m_1\sim\mathcal{O}(\text{TeV})$, we take $m_1\gg m_S\rightarrow0$ and $\sigma_0$ in the non-resonant region. For example, with $m_{1}\approx5$ TeV WIMP with freeze-out Yukawa couplings $\delta\approx0.2$, $\rho_1=0.08$, and $\mu=5~\text{GeV}$, $Y_{1}$ freezes out as a cold relic with $T_{\rm f.o.}\approx\frac{m_1}{16.5}\approx303~\text{GeV}$. Its comoving density $Y_{Y_1}\equiv\frac{n_{Y_{1}}}{s}$ at the time of freezeout is given by \cite{kolbturner}
\begin{equation}\label{eq:fo}
Y_{Y_1,\text{f.o.}}=\frac{7.58g_{*}^{1/2}x_{f.o.}^{2}}{g_{*S}M_{\text{Pl}}m_{1}\sigma_{0}},
\end{equation}
where $g_{*S}$ is the effective number of degrees of freedom in entropy. Note that if $Y_1$ does not decay, its would-be relic abundance today $Y_{Y_1}^{\tau\rightarrow\infty}\approx Y_{Y_{1,\text{f.o.}}}$ 

Following the schematic illustration in Fig.~\ref{fig:cartoon}, we expect the observed abundances of DM and baryons to be proportional to the freeze out abundance found in Eq. \ref{eq:fo}. 
\subsection{C and CP Violation}\label{sec:cpv}
\begin{figure} 
\centering
\includegraphics[width=0.9\textwidth, trim= 0 300 0 300, clip]{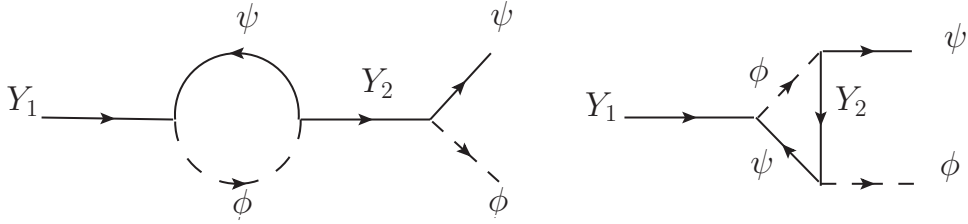}
\caption{Loop diagrams interfere with the tree-level diagram to produce a nonzero asymmetry between $Y_{1}$ decays to $\phi/\psi$ and $\phi^{*}/\bar{\psi}$}
\label{fig:cpdiagrams}
\end{figure}
C- and CP-violation are achieved by the decay of the Majorana fermions $Y_1$ following their freeze out. The CP asymmetry arising from $Y_{1}$ decays is defined as 
\begin{equation}\label{eq:cpasym}
\epsilon_{1}=\frac{\Gamma(Y_{1}\rightarrow\phi\bar{\psi})-\Gamma(Y_{1}\rightarrow\phi^{*}\psi)}{\Gamma(Y_{1}\rightarrow\phi\bar{\psi})+\Gamma(Y_{1}\rightarrow\phi^{*}\psi)}
\end{equation}
The denominator of Eq.~\ref{eq:cpasym} can be approximated as twice the tree-level decay rate, $\Gamma_{0}(Y_{1}\rightarrow\phi\bar{\psi})$. For complex WIMP Yukawa couplings, interference between the tree-level and loop-level Feynman diagrams shown in Fig. \ref{fig:cpdiagrams} gives rise to a non-vanishing numerator in Eq. \ref{eq:cpasym}. Although in analogy $Y_2$ decay may generate a CP asymmetry as well, its contribution to the DM/baryon asymmetry is generally washed out with $m_2>m_1$ and $|\eta_1|\ll|\eta_2|$ (leading to sizable $\epsilon_{1}$ but in-equilibrium decay of $Y_2$).

In many baryogenesis models based on massive particle decay, the decay products are much lighter than the decaying particle and thus can be approximately taken as massless. For WIMP cogenesis we include full mass-dependence since WIMP freezeout generically requires $m_{1}\sim\mathcal{O}(100)~\text{GeV}-10~\text{TeV}$ while the intermediate decay products $\phi$ and $\psi$ are experimentally constrained to have masses $\gtrsim O(100)~\text{GeV}-O(\text{TeV})$ (Sec.~\ref{sec:collider}). 

With this in mind and using the Optical Theorem, we find the CP-asymmetry:
\begin{eqnarray}\label{eq:cpfull}
\epsilon_{1}&=&-3\frac{\sqrt{xa}\text{Im}\big[(\eta_2^*\eta_1)^2\big]}{8\pi|\eta_1|^2b^2}\bigg\{\frac{1}{1-x}+b^2+c^2\ln\bigg[\frac{c^2+\frac{b^2}{2a}(1-a)}{c^2+\frac{b^2}{2a}(1+a)}\bigg] \bigg\}
\end{eqnarray}
where $a\equiv(1+\frac{m_\psi^2-m_\phi^2}{m_1^2})^{-2}\big[(1-\frac{m_\psi+m_\phi^2}{m_1^2})^2-\frac{4m_\psi^2m_\phi^2}{m_1^4}\big]$, $b^2\equiv a\big(1+\frac{m_\psi^2-m_\phi^2}{m_1^2}\big)^2$, $c^2\equiv\frac{2m_{\phi}^{2}-m_{1}^{2}-m_{2}^{2}}{m_{1}^{2}}$ and $x=\frac{m_{2}^{2}}{m_{1}^{2}}$. The factor of 3 represents the color multiplicity. Note this is the contribution to the CP-asymmetry of $Y_{1}$ decays to a single generation. To simplify our analyses, we assume the three flavors of $\phi$ and $\psi$ are (nearly) degenerate in mass. Under this assumption, there is an additional multiplicative factor of 3 to account for the contributions $Y_{1}$ decays to the all flavors. Also note that Eq.~\ref{eq:cpfull} reproduces the familiar CP-asymmetry result for leptogenesis \cite{cpviolation} in the limit of $m_{1}\gg m_{\psi,\phi}$. The above expression shows how the asymmetry is intimately tied to the mass and couplings of the $Y_{1},~m_{\phi},~\text{and}~m_{\psi}$. In Section \ref{sec:quarkresults}, we show contours of constant $\Omega_{DM}$ in the $(m_{1},\rho)$ plane with $\epsilon_{1}$ taking the form of Eq. \ref{eq:cpfull}.

\subsection{Generalized Baryon Number Conservation and Generation of Asymmetries}\label{sec:bv} 
In order for a matter asymmetry to be produced, the corresponding baryon or DM number must be violated by the interactions in the model. In this model both SM baryon number and DM number are violated in the last stage of the decay chain as illustrated in Fig.~\ref{fig:quarksteps}. Nevertheless a generalized baryon number $G={B+2X}$ is conserved (remains 0 assuming no pre-existing asymmetry) thanks to the ADM $\chi$ and baryonic matter sharing interactions through intermediate states $\phi, \psi$. 

The CP-violation in $Y_1$ decay (Section \ref{sec:cpv}) produce an asymmetry between intermediate states (i.e., baryon/ADM parents), $\phi$ and $\psi$ and their conjugates, which is inherited by their decay products, $\chi$ and $udd$, and ultimately becomes the source of all (asymmetric) matter today. The changes in the generalized baryon number for each decay process are given by:
\begin{eqnarray}
\Delta G_{Y_1\rightarrow\phi\psi}=G_{\phi}+G_{\psi}-G_{Y_1} \label{eq:3}\\
\Delta G_{\phi\rightarrow\chi d}=1/3 +G_{\chi}-G_{\phi} \label{eq:4}\\
\Delta G_{\psi\rightarrow\phi u}=1/3+G_{\phi}-G_{\psi}\label{eq:5},
\end{eqnarray}
where we have used the fact that for quarks $G_q=B_q=1/3$. Furthermore, due to the Majorana nature of $Y_1$, its natural $U(1)_{B+2X}$ charge is $G_{Y_1}=0$. Then requiring all the above interactions to conserve $G$, we may obtain the solutions for the charge assignments: $G_{\chi}=-1/2$, $G_{\phi}=-G_{\psi}=-1/6$, as listed in Table.~\ref{tab:table1}. The net result of the decay chain is $udd+\chi\chi$, violating the SM baryon number and DM number by 1 and 2 units respectively, while the net generalized baryon number $G$ is conserved.
So the generalized baryonic charge carried by the ADM density cancels that of a baryon asymmetry density and the universe has trivial net generalized baryon number. 
\subsection{WIMP Decays and Production of Matter Asymmetries}\label{sec:deq}
We consider the asymmetry grandparent, $Y_1$, decays well after its freezeout but before BBN, i.e., $1 {\rm MeV} \lesssim T_{Y_1, \rm dec} \lesssim T_{\rm f.o.}$, so that we can treat the freezeout and decay-triggered cogenesis as nearly decoupled processes and retain the conventional success of BBN. The $Y_{1}$ decay rate at $T<m_{1}$ is $\Gamma_{Y_{1},\rm dec}\approx\frac{|\eta_{1}|^{2}m_{1}}{8\pi}$. Following Eq. \ref{eq:fotemp}, the freezeout occurs around the temperature $T_{\text{f.o.}}\sim200-300$ GeV for TeV-scale mass $Y_{1}$. The requirement that it decay between freezeout and BBN gives the range of allowed decay couplings: $10^{-15}\lesssim|\eta_{1}|\lesssim10^{-9}$. For simplicity we assume the subsequent SM $B$- and DM $\chi$-number violating decay of $\phi, \psi$ to $udd,\chi$ are prompt relative to $H$, i.e., in equilibrium, so that the matter asymmetries are immediately distributed upon $Y_1$ decay. This assumption also simplifies the Boltzmann equations, since $n_\psi, n_\phi$ can be set as equilibrium distribution.

With $Y_{1}$ freezeout occurring well before its decay, the late-time evolution of comoving density $Y_{Y_1}$ satisfies the following Boltzmann equation for a decaying species: 
$$\frac{dY_{Y_1}}{dx}=\frac{-x\langle\Gamma(Y_1\rightarrow\phi\psi)\rangle}{2H(m_{1})}(Y_{Y_1}-Y_{Y_1}^{\text{eq}})$$ where $x=m_{1}/T$ and $H(m_{1})=H(T=m_{1})$. The initial condition for $Y_{Y_1}$ of this stage of evolution is set by the would-be abundance of $Y_1$ after its freezeout: $Y_{Y_1}(0)\approx Y_{Y_1, \rm f.o.}$ where $Y_{Y_{1},\text{f.o.}}$ is given in Eq. \ref{eq:fo}. 

We now write down the Boltzmann equations governing the evolution of $\phi,~\psi$ number densities. This evolution is determined by three processes: CP-violating $Y_{1}$ decays and their inverse, $Y_{1}$ mediated $\phi/\psi$ scattering to their conjugates (and vice versa), and CP-conserving $\phi/\psi$ (as well as their conjugates) decays. 

For convenient notations, we define the generalized baryon number density $n_{G}$ which is the sum of $\phi/\psi$ asymmetries:
\begin{equation}\label{eq:genmatter}
n_{G}=\frac{n_{\phi}-n_{\phi^{*}}}{2}+\frac{n_{\psi}-n_{\bar{\psi}}}{2}
\end{equation}

Once simplified, the $\phi$ asymmetry, $n_{\phi}-n_{\phi^{*}}\equiv n_{\Delta\phi}$ evolves according to 
\begin{eqnarray}\label{eq:phiev}
\dot{n}_{\Delta\phi}+3Hn_{\Delta\phi}&=& \epsilon_{1}\langle\Gamma(Y_{1}\rightarrow\phi\psi)\rangle(n_{Y_{1}}-n_{Y_{1}}^{\text{eq}}-\frac{Y_{G}}{2\epsilon_{1}}n_{Y_{1}}^{\text{eq}})-2n_{G} n_{\gamma}\langle\sigma(\phi\psi\rightarrow\phi^{*}\bar{\psi})|\vec{v}|\rangle\nonumber\\&-&\langle\Gamma(\phi\rightarrow\chi+d)\rangle\big[(n_{\phi}-n_{\phi}^{\text{eq}})-(n_{\phi^{*}}-n_{\phi^{*}}^{\text{eq}})\big]\nonumber\\&+&\langle\Gamma(\psi\rightarrow\phi+u)\rangle\big[(n_{\psi}-n_{\psi}^{\text{eq}})-(n_{\bar{\psi}}-n_{\bar{\psi}}^{\text{eq}})\big],
\end{eqnarray}
where $\epsilon_{1}$ is the CP asymmetry given in Eq. \ref{eq:cpfull}, $\langle\Gamma\rangle$'s are thermally averaged decay rates, $Y_{G}\equiv\ n_{G}/s=\frac{1}{2s}[(n_{\phi}-n_{\phi^{*}})+(n_{\psi}-n_{\bar{\psi}})]$, and $n_{\gamma}$ is the photon radiation density. The equation governing the cosmological evolution of the $\psi$ asymmetry is
\begin{eqnarray}\label{eq:psiev}
\dot{n}_{\Delta\psi}+3Hn_{\Delta\psi}&=& \epsilon_{1}\langle\Gamma(Y_{1}\rightarrow\phi\psi)\rangle(n_{Y_{1}}-n_{Y_{1}}^{\text{eq}}-\frac{Y_{G}}{2\epsilon_{1}}n_{Y_{1}}^{\text{eq}})-2n_{G} n_{\gamma}\langle\sigma(\phi\psi\rightarrow\phi^{*}\bar{\psi})|\vec{v}|\rangle\nonumber\\ &-&\langle\Gamma(\psi\rightarrow\phi+u)\rangle\big[(n_{\psi}-n_{\psi}^{\text{eq}})-(n_{\bar{\psi}}-n_{\bar{\psi}}^{\text{eq}})\big]
\end{eqnarray}
We can see that the main difference between the $\phi$ and $\psi$ Boltzmann evolution is that the term governing $\psi$ decays changes sign and there is no term for $\psi$-number increasing $\phi$ decays. Note that in these evolution eqs., the terms proportional to $Y_{G}$ can potentially wash out the produced asymmetries (inverse decay of $Y_1$ and the 2-2 scattering). Assuming prompt $\phi, \psi$ decays, we set $n_{\phi}=n_{\phi}^{\text{eq}}, n_{\psi}=n_{\psi}^{\text{eq}}$, such that the contribution from these decays vanish. Additional potential washout processes of $udd\chi\chi\rightarrow Y_{1}$ and $udd\chi\chi\rightarrow \bar{u}\bar{u}\bar{d}\bar{\chi}\bar{\chi}$ are negligible owing not only to Boltzmann suppression, but also to the high dimension of the effective operators responsible for these processes.

Based on Fig.~\ref{fig:quarksteps} and our earlier discussion, upon decays of $\phi$ and $\psi$, $n_G$ or $Y_G$ leads to baryon asymmetry density $n_B$ and DM asymmetry density $n_\chi$ with the robust relation: 
\begin{equation}\label{eq:numdens}
n_G=(n_{\Delta\phi}+n_{\Delta\psi})/2=n_B=n_\chi/2
\end{equation}

The general solution of the Boltzmann equations gives the comoving generalized matter asymmetry $Y_G$ today:
\begin{eqnarray}\label{eq:fullsol}
Y_{G}(0)&=&\epsilon_{1}\int_{0}^{T_{dec}}\frac{dY_{Y_{1}}}{dT}\exp\Big(-\int_{0}^{T}\frac{\Gamma_{W}(T')}{H(T')}\frac{dT'}{T'}\Big)dT\nonumber\\&+&Y_{B}^{\text{initial}}\exp\Big(-\int_{0}^{T_{\text{initial}}}\frac{\Gamma_{W}(T)}{H(T)}\frac{dT}{T}\Big)
\end{eqnarray}
where $\Gamma_{W}$ is the rate of processes washing out the asymmetry. Assuming that there is no primordial asymmetry before WIMP cogenesis occurs, $Y_{B}^{\text{initial}}=0$. Taking our simplifying assumption that $Y_1$ decays well after its freeze out, we automatically work in the weak washout regime and drop the exponential factor in Eq. \ref{eq:fullsol}. This yields a robust solution depending solely on the would-be WIMP miracle abundance of $Y_1$ and the CP asymmetry $\epsilon_1$:
\begin{equation}\label{eq:genasymp}
Y_B(\infty)=Y_\chi(\infty)/2=Y_G(\infty)\approx \epsilon_{1} Y_{Y_1, \text{f.o.}}
\end{equation}
Provided efficient annihilation that depletes the symmetric component of $\chi$, the above asymptotic solution of $n_B, n_\chi$ give rise to the baryon and DM abundances today:
\bea\label{eq:endens}
\Omega_{\chi}(\infty)&=&\frac{2m_{\chi}s_{0}}{\rho_{c}}\epsilon_{1}Y_{Y_1,\text{f.o.}}\\
\Omega_{B}(\infty)&=&\frac{c_sm_{n}s_{0}}{\rho_{c}}\epsilon_{1}Y_{Y_1,\text{f.o.}},
\eea
where $s_{0}=2970~\text{cm}^{-3}$ is the radiation entropy density today and $\rho_{c}=3H_{0}^{2}/8\pi G\approx3.5\times10^{-47}~\text{GeV}^{4}$ is the critical energy density, $m_n\approx 1$ GeV is the SM baryon mass. $\epsilon_{1}$, $Y_{Y_1,\text{f.o.}}$ have been calculated in earlier sections. Based on the discussion about $c_s$ and Eq. \ref{eq:admmass} in Sec.~\ref{sec:intro} , the observed relation $\Omega_{DM}\approx5~\Omega_{B}$ fixes $m_{\chi}=2.5~\text{GeV~or}~m_{\chi}=0.72-0.89~\text{GeV}$ for $Y_1$ decay after or before EWPT, respectively.

\subsection{Numerical Results}\label{sec:quarkresults}
We now scan parameter space to demonstrate viable regions that predicts $\Omega_{\chi}=\Omega_{\rm DM}\approx5\Omega_B$ as observed. The relevant parameters includes the masses $(m_{1},~m_{2},~m_{\phi},~m_{\psi},~m_{\chi},~m_S)$ and couplings $(\eta_{1},~\eta_{2},~\rho,~\delta,~\mu)$. We take $\eta_{1}$ to be real such that the CP-asymmetry in Eq. \ref{eq:cpfull} can be written in terms of a complex phase of $\eta_{2}$: $\text{Im}[(\eta_{1}^{*}\eta_{2})^{2}]\rightarrow|\eta_{1}|^{2}|\eta_{2}|^{2}\sin(2\theta_{2})$. In our analyses, we fix $\theta_{2}=\pi/4$ to bound the CP-asymmetry from above. 

Due to the color charges of $\phi$ and $\psi$, their masses are effectively constrained by collider experiments (see Section \ref{sec:collider}). This immediately constrains the mass of the lighter of the Majorana fermion $m_{1}\gtrsim3~\text{TeV}$ such that $Y_{1}\rightarrow\phi\psi$ remains kinematically open for $m_{\psi}\gtrsim m_{\phi}\sim\text{TeV}$. The symmetric component of ADM is efficiently depleted through annihilations to the hidden sector, e.g. $\chi\bar{\chi}\rightarrow SS$, which requires $m_{\chi}>m_{S}$ such that the annihilation process is kinematically open. 
\begin{figure}[h!]
\centering
\captionsetup{singlelinecheck = false, justification=raggedright, font=normalsize}
\includegraphics[width=0.7\linewidth, trim=110 210 110 210, clip]{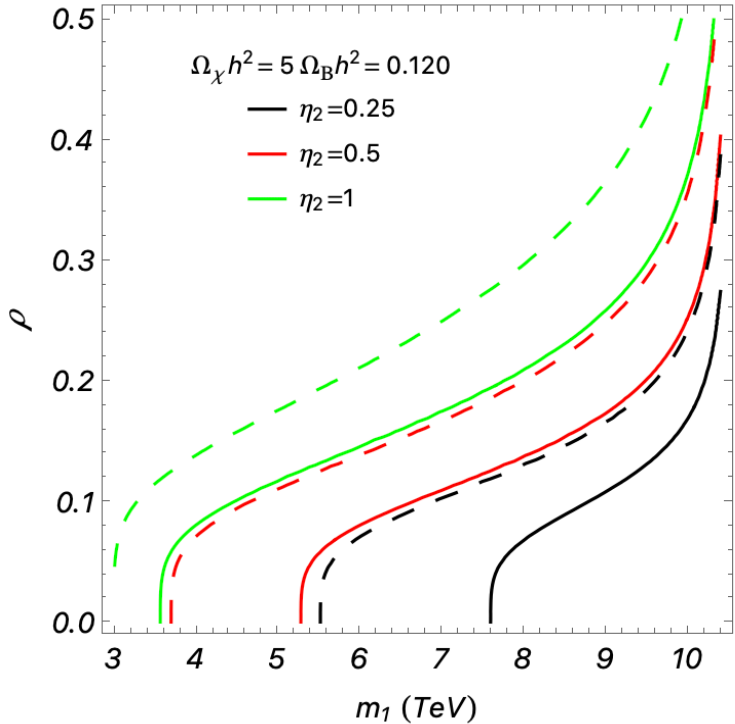} 
\caption{Contours of $\Omega_{\chi}, \Omega_B$ as a function of Yukawa coupling $\rho_1$ and $Y_{1}$ mass $m_{1}$ for different values of $\eta_{2}$ in WIMP cogenesis for baryons. The solid (dashed) lines correspond to the case where asymmetries in DM and baryons are produced before (after) the EWPT with $m_{\chi}=0.72~\text{GeV}$ ($m_{\chi}=2.5~\text{GeV}$). The benchmark parameters used are: $\delta=0.2$, $\mu=5~\text{GeV}$, $m_{S}\rightarrow0~\text{GeV}$, $m_{\phi}=1.2~\text{TeV}$, $m_{\psi}=1.7~\text{TeV}$, and $m_{2}=10.5~\text{TeV}$.}
\label{fig:quarkplot1}
\end{figure}

Taking benchmark values of $m_{2}\gtrsim10~\text{TeV},~m_{\phi}\approx1.2~\text{TeV},~m_{\psi}\approx1.7~\text{TeV},~\delta=0.2,~\text{and}~\mu\approx5~\text{GeV}$, with \textit{Mathematica} \cite{Mathematica} we plot contours of $\Omega_{DM} h^2=0.120\pm0.001$, $\Omega_B h^2=0.0224\pm0.0001$ \cite{planck} in the $(m_{1}, \rho_1)$ plane, as shown in Fig. \ref{fig:quarkplot1}. Because the baryon asymmetry is directly produced by $Y_{1}$ decays, it may be produced before or after the EWPT. Comparing the case of $Y_1$ decay before vs. after EWPT, we see that a smaller $Y_1$ Yukawa coupling to $S$ and larger $m_1$ (for a given $|\eta_2|$) are required to produce the observed DM abundance when the asymmetry is produced before EWPT due to the sphaleron's moderate washout of the SM baryon asymmetry. This is because $Y_{Y_1, \text{f.o.}}\sim1/\sigma_{0}\sim m_1^2/\rho_1^4$ giving a larger decaying $Y_1$ abundance to compensate for this washout. The CP-asymmetry produced by $Y_{1}$ decays, as given in Eq. \ref{eq:cpfull}, must be sufficient to produce the observed abundances of DM and baryons for $\rho\sim0.1$ (as discussed in Sec. \ref{sec:model}) and $m_{1}\sim\mathcal{O}~(\text{TeV})$. To give an example, with $Y_{1}$ Yukawa coupling $\eta_{2}=0.5$, $m_{2}=10.5~\text{TeV}$, $m_{1}=5~\text{TeV}$, $m_{\psi}=1.7~\text{TeV}$, and $m_{\phi}=1.2~\text{TeV}$ the CP-asymmetry is $|\epsilon_{1}|\approx1\%$. 

\section{WIMP Decay to Leptons and ADM}\label{sec:lepton}

In the following section, we present a WIMP cogenesis model that directly produces a lepton asymmetry. As with other models of leptogenesis, the asymmetry must be produced before EWPT such that sphalerons may transfer the lepton asymmetry into the observed baryon asymmetry. Here, we introduce the fields, interactions, and discuss the differences from WIMP cogenesis with baryons presented in the last section.

\subsection{Model Setup}

\begin{figure} 
\centering
\includegraphics[width=0.35\textwidth, trim=15 100 20 125, clip]{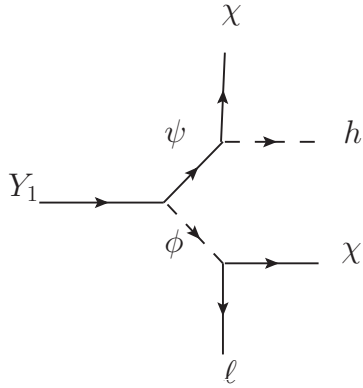}
\caption{Feynmann diagram of the decay chain for WIMP cogenesis with leptons. $\psi$ may also decay to electroweak gauge bosons $Z$ and $W^\pm$}
\label{fig:leptonsteps}
\end{figure}

The first two stages of WIMP cogenesis with leptons are identical to the model discussed above: the Majorana fermion, $Y_{1}$, undergoes freezeout via Yukawa and cubic interactions with singlet scalar $S$ followed by out-of-equilibrium and CP-violating decays to (unstable) intermediate states $\phi$ and $\psi$. Again, the Majorana fermion, $Y_1$, is a SM gauge singlet, but now the intermediate states are charged under SM $SU(2)_{L}\times U(1)_{Y}$, such that the decays $\psi^0\rightarrow\chi h,~\chi Z$, $\psi^{\pm}\rightarrow\chi W^\pm$ and $\phi\rightarrow\chi \ell$ are possible, where $h, W^\pm, Z, \ell$ are the SM Higgs, electroweak gauge bosons, and left-handed leptons, respectively. The Lagrangian is identical to that in Eqs. \ref{eq:Lyuk} and \ref{eq:Ldark} up to modification of the Yukawa interactions: 
\begin{eqnarray}\label{eq:leptoL}
\mathcal{L}_{\text{Yukawa}}\rightarrow-\alpha_{ijk}\phi_{i}\bar{L}_{i}\chi_k^{c}-\beta_{ii} H\bar{\psi}_i\chi_i
\end{eqnarray}
where $L$ is the left-handed lepton doublet, $H$ is the Higgs doublet, $i=1, 2, 3$ is flavor indices, and $\alpha_{ijk}$ is antisymmetric in flavor indices. Note that this model possesses a $U(3)$ flavor symmetry which prevents new sources of FCNC. As discussed in Sec. \ref{sec:quark} the U(3) symmetry is optional provided that $10^{-7}\lesssim\alpha\lesssim0.1$, while the DM direct detection signal may be absent with such small couplings. The charge assignments are summarized in Table \ref{tab:table2}. A $Z_4$ symmetry is imposed to ensure DM stability and prevent $Y_1$ decay through $Y_1LH$ portal. The decay chain is illustrated in Fig.\ref{fig:leptonsteps}. The CP asymmetry is generated by the same process as illustrated in Fig.~\ref{fig:cpdiagrams}.
\begin{table}[h!]
\centering
\begin{tabular}{|l|l|l|l|l|l|l|}
\hline
        & $SU(3)_C$      & $SU(2)_L$ & $U(1)_Y$  & $U(1)_{L+2X}$ & $Z_{4}$ \\ \hline
$Y_{1,2}$ & $\bf{1}$       & $\bf{1}$  & $~~0$           & $~~0$              & $-1$              \\ \hline
$\psi$  & $\bf{1}$ & $\bf{\bar{2}}$  & $+1$         & $-1/2$            & $+i$              \\ \hline
$\phi$  & $\bf{1}$       & $\bf{2}$  & $-1$        & $+1/2$           & $+i$              \\ \hline
$\chi$  & $\bf{1}$       & $\bf{1}$  & $~~0$         & $-1/2$           & $+i$              \\ \hline
$S$     & $\bf{1}$& $\bf{1}$  & $~~0$        & $~~0$            & $+1$              \\ \hline
$L$     & $\bf{1}$       & $\bf{2}$  & $-1$        & $+1$            & $+1$              \\ \hline
$H$     & $\bf{1}$& $\bf{\bar{2}}$  & $+1$       & $~~0$            & $+1$              \\ \hline
\end{tabular}
\caption{Quantum numbers of the relevant particles in WIMP cogenesis with leptons.}
\label{tab:table2}
\end{table}

In analogy to WIMP cogenesis with baryons, the shared interactions through intermediate $\phi, \psi$ permit a generalized global lepton number symmetry $U(1)_{L+2X}$ with conserved charge $G'$. The corresponding charge assignment is: $G'_{\chi}=1/2$, $G'_{Y_1}=0$, $G'_{\phi}=1/2$, and $G'_{\psi}=-1/2$. As shown in Fig.~\ref{fig:leptonsteps}, the second stage of of the decay chain violates SM lepton and DM number, giving rise to 1 unit of $L$-number and 2 units of $X$-number. After all the decays have taken place, efficient annihilations deplete the symmetric components of ADM and leptons, leaving an abundance of $\chi$ and $L$. A key difference from the model in Sec.~\ref{sec:quark} is that the asymmetry must be produced before EWPT such that sphalerons convert the lepton asymmetry into the observed baryon asymmetry, i.e., $T_{\text{f.o}}>T_{\text{dec}}\gtrsim T_{\text{EWPT}}$. As in the quark model, $\chi\bar{\chi}$ depletion occurs through annihilation to $S$. This depletion may also receive contributions from $\phi$-mediated annihilation to leptons, due to the weaker constraints on ADM-lepton couplings (relative to ADM-quark couplings) \cite{admeffectiveops}.

\begin{figure}[h!]
\centering
 \includegraphics[width=0.9\textwidth, trim=0 275 0 275, clip]{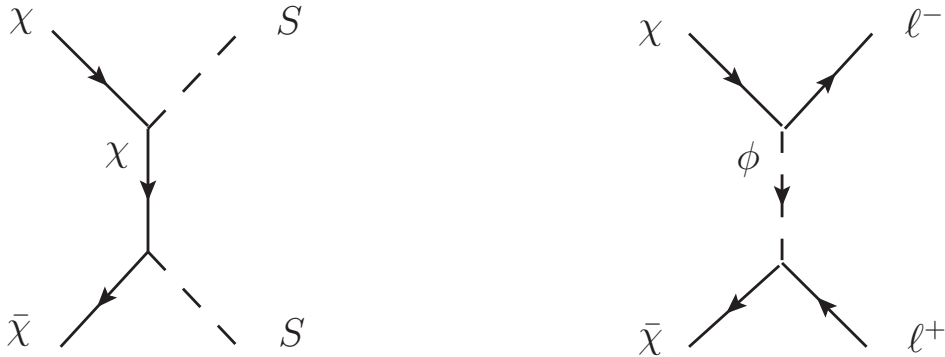}
  \caption{Diagrams contributing to $\chi\bar{\chi}$ depletion.} 
  \label{fig:dmannlepto}
\end{figure}

We can then apply most results from Sections \ref{sec:freezeout}-\ref{sec:deq} by analogy, with some modifications. The most straightforward change is the dropping of the color factor in the CP-asymmetry of Eq. \ref{eq:cpfull}. More subtle is the change to the DM mass prediction. Due to the different Yukawa interactions, the prediction of the relation $c_{s}=\frac{n_{B}}{n_{B-L}}$ in this model differs from that in the WIMP cogenesis with baryons. In addition, as noted, WIMP cogenesis with leptons needs to occur before EWPT when sphaleron processes are active. The limits of interest are the same as those detailed in the previous section. The solutions in these two limits are (see Appendix \ref{sec:cs2}) 
\begin{eqnarray}\label{eq:leptocs}
c_{s}=\frac{n_{B}}{n_{B-L}}=\begin{cases} \frac{8N_{f}+4N_{H}}{30N_{f}+13N_{H}} & m_{\phi,\psi}\ll T_{\text{EWPT}}\\
\frac{8N_{f}+4N_{H}}{22N_{f}+4N_{H}} & m_{\phi,\psi}\gg T_{\text{EWPT}}
\end{cases}
\end{eqnarray}
where $N_F$ and $N_H$ are again the number of generations of fermions and Higgs, respectively. With $N_{f}\rightarrow3$ and $N_{H}\rightarrow1$ in Eq. \ref{eq:leptocs} with gives $c_{s}=28/103$ for $m_{\phi,\psi}\ll T_{\text{EWPT}}$ or $c_{s}=28/79$ for $m_{\phi,\psi}\gg T_{\text{EWPT}}$. All together, the relation between lepton, baryon, and ADM comoving densities is akin to Eq. \ref{eq:genasymp}: $Y_{L}=Y_{\chi}/2=\frac{|c_{s}-1|}{c_{s}}Y_{B}$.
Following the same procedure as Sec.~\ref{sec:deq}, in the weak washout regime we obtain ADM abundance with the same form as Eq. \ref{eq:endens}:
\bea\label{eq:leptoendens}
\Omega_{\chi}(\infty)&=&\frac{2m_{\chi}s_{0}}{\rho_{c}}\epsilon_{1}Y_{Y_1,\text{f.o.}}\\
\Omega_{B}(\infty)&=&\frac{c_{s}m_{n}s_{0}}{|c_{s}-1|\rho_{c}}\epsilon_{1}Y_{Y_1,\text{f.o.}}.
\eea

The observed ratio $\Omega_{DM}/\Omega_{B}\approx5$ fixes the mass of the ADM candidate $m_{\chi}=\frac{5c_s}{2|c_{s}-1|}m_{n}$. With the values for $c_{s}$ given in Eq. \ref{eq:leptocs}, the range of $\chi$ masses is $0.93-1.37$~\text{GeV}. 

\subsection{Numerical Results}\label{sec:leptonresults}
We now scan model parameters to find viable region giving the observed matter abundances. The relevant parameters includes the masses $(m_{1},~m_{2},~m_{\phi},~m_{\psi},~m_{\chi},~m_S)$ and couplings $(\eta_{1},~\eta_{2},~\rho,~\delta,~\mu)$. We also take the same parametrization for the CP-asymmetry relevant Yukawa couplings, $\eta_1$ and $\eta_2$, as in Sec. \ref{sec:quarkresults}. 

The constraints arising from colliders on exotic electroweak states ($\phi$ and $\psi$ in this model) are less stringent than those on exotic colored states, allowing us to explore sub-TeV masses for $\phi$, $\psi$, and even the grandparent, $Y_{1}$. There is a caveat to this: if the mass of the decaying WIMP is too light, it freezes out \textit{after} the EWPT, thus its lepton asymmetry producing decays would occur when sphaleron processes, necessary for the conversion into the observed baryon asymmetry, are no longer effective. For $Y_{1}$ decays to happen after freezeout, but before EWPT, we require $100~\text{GeV}\lesssim T_{Y,\text{dec}}\lesssim T_{\text{f.o}}$. With a $m_{1}\sim1~\text{TeV}$, the freezeout occurs at or just after EWPT, according to Eq. \ref{eq:fotemp}. 
\begin{figure}[h!]
\centering
\captionsetup{singlelinecheck = false, justification=raggedright, font=normalsize}
\includegraphics[width=0.7\linewidth, trim=110 215 110 220, clip]{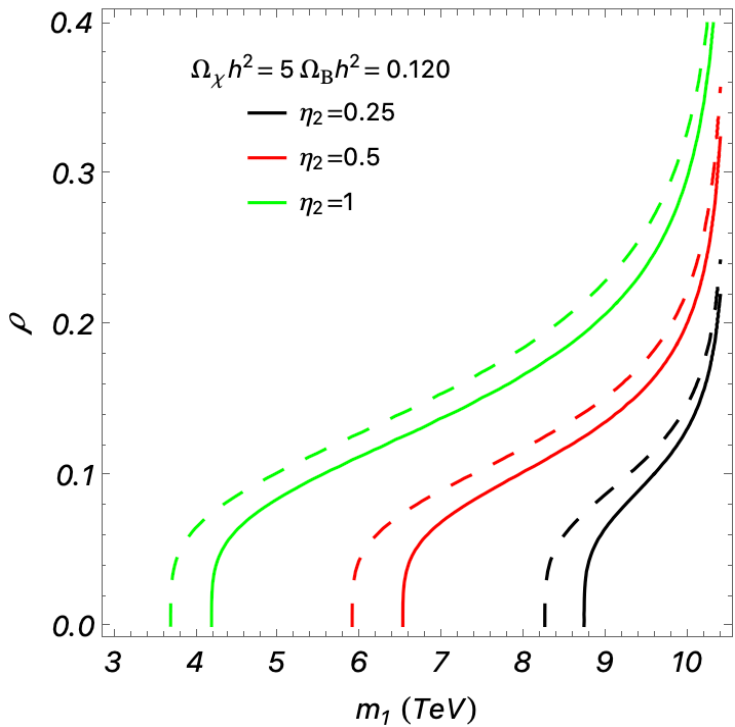} 
\caption{Contours of $\Omega_{\chi}, \Omega_B$ as a function of Yukawa coupling $\rho_1$ and $Y_{1}$ mass $m_{1}$ for different values of $\eta_{2}$ for WIMP cogenesis with leptons. The solid lines correspond to $m_{\chi}=0.93~\text{GeV}$ and dashed lines to $m_{\chi}=1.37~\text{GeV}$, both cases with with $S$ mass $m_{S}=0~\text{GeV}$. Other benchmark parameters are: $\delta=0.2$, $\mu=5~\text{GeV}$, $m_{\phi}=700~\text{GeV}$, $m_{\psi}=740~\text{GeV}$, and $m_{2}=10.5~\text{TeV}$.}
\label{fig:leptonplot1}
\end{figure}

Since $\psi$ contributes to the matter asymmetry via $\psi^0\rightarrow\chi h, \chi Z,~\text{and/or}~\psi^{\pm}\rightarrow\chi W^{\pm}$ it requires $m_{\psi}$ greater than at least $m_W$. Similarly, since $\phi$ decays to $\mathcal{O}(\text{GeV})$ mass $\chi$ and SM leptons, $m_{\phi}\gtrsim\mathcal{O}(\text{GeV})$ is required. That said, collider constraints (see \ref{sec:collider}) on new electroweak states require these states be much heavier than the above kinematic requirements. Fig. \ref{fig:leptonplot1} shows the DM abundance as a function of $Y_1$-S Yukawa coupling and $m_{1}$, in the range of $1~\text{TeV}<m_{1}<10~\text{TeV}$. In these numerical analyses, we take the functional form of Eq.~\ref{eq:cpfull} and Eq. \ref{eq:fo} for the CP-asymmetry and freezeout abundance of $Y_{1}$, respectively. 

As can be seen in Fig. \ref{fig:leptonplot1}, a smaller coupling $\rho_1$ is required in the case with smaller $m_\chi$. This is related to whether interactions of $\psi,\phi$ Yukawa and gauge interactions contribute to chemical equilibration along with sphalerons. When they contribute ($m_{\phi,\psi}\ll T_{\text{EWPT}}$), there is further washout of the produced asymmetry. A smaller Yukawa coupling $\rho$ compensates for this washout since $Y_{\infty}\propto\sigma^{-1}$. 

\section{Phenomenology and Constraints}\label{sec:pheno}

\subsection{Collider Phenomenology}\label{sec:collider}
\textbf{WIMP Decay to Baryons and ADM} (Sec.~\ref{sec:quark})
\begin{figure}[h!]
\centering
 \includegraphics[width=0.8\linewidth, trim=0 240 0 250, clip]{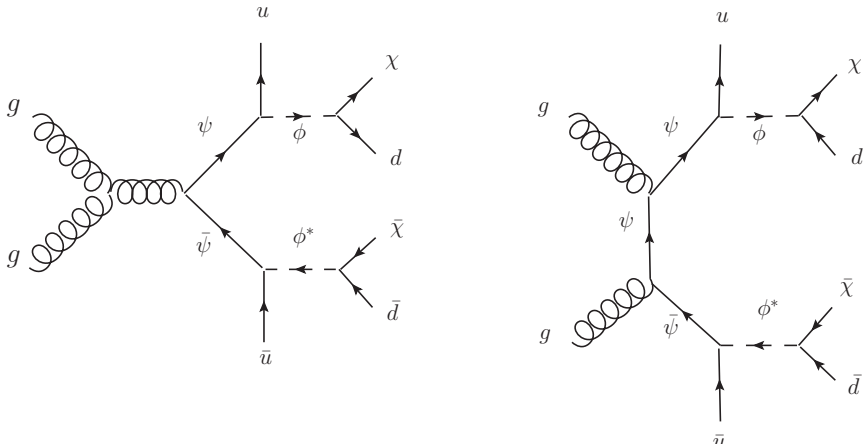}
 \caption{Diagrams relevant for $\psi$ searches at hadron colliders (WIMP cogenesis with baryons).} 
 \label{fig:quarkcolliderpsi}
\end{figure}

\begin{figure}[h!]
\centering
 \includegraphics[width=0.7\linewidth, trim=0 250 0 250, clip]{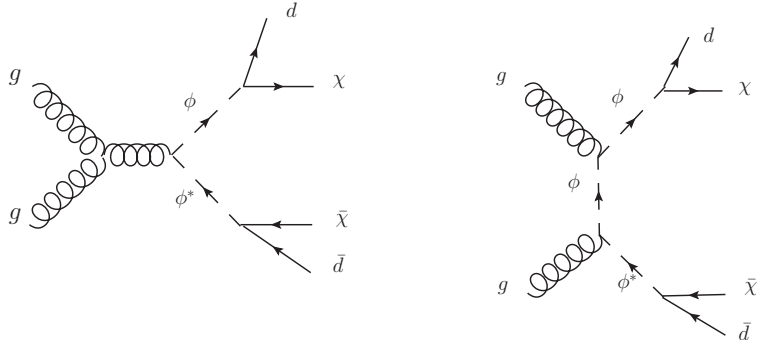}
  \caption{Diagrams relevant for $\phi$ searches at hadron colliders (WIMP cogenesis with baryons). }
  \label{fig:quarkcolliderphi}
\end{figure}

In the model where the WIMP decays to quarks (Sec. \ref{sec:quark}), SM charged colored scalars and fermions, $\phi_{i}$ and $\psi_{i}$ respectively, are introduced. Owing to the color charges carried by these intermediate states, the LHC bounds on their masses are strong. As outlined in Sec. \ref{sec:quark}, $\psi$ decays through intermediate scalar $\phi$ to 2 SM quarks and singlet ADM candidate $\chi$, and $\phi$ decays to an SM quark and $\chi$. These states are pair-produced at the LHC dominantly through gluon fusion, with subsequent decays $\phi\rightarrow j+\slashed{E}_{T}$, $\psi\rightarrow jj+\slashed{E}_{T}$, rendering typical signatures: $pp\rightarrow\psi\bar{\psi}\rightarrow4j+\slashed{E}_{T}$ and $pp\rightarrow\phi\phi^{*}\rightarrow jj+\slashed{E}_{T}$. The relevant diagrams are shown in Figs. \ref{fig:quarkcolliderpsi} and \ref{fig:quarkcolliderphi}. 

LHC searches for squarks, $\tilde{q}$, and gluinos, $\tilde{g}$, in the presence of neutralino LSP $\tilde{\chi}_{1}^{0}$ are relevant for constraining the masses of $\phi$ and $\psi$ in our model. In particular the bound in the massless LSP limit applies since the corresponding particle in WIMP cogenesis, $\chi$ has a mass of $\mathcal{O}(\text{GeV})$, significantly smaller than those of $\phi$ and $\psi$. Specifically, both $\psi$ and $\tilde{g}$ decay to $jj+\slashed{E}_{T}$ via intermediate colored scalars with production cross sections differing only by a group theory factor, for which we correct. Simplified model searches at $13~\text{TeV}$ from CMS with $137~\text{fb}^{-1}$ of data place bounds on the gluino mass in the presence of a massless LSP, neutralino $\tilde{\chi}_{1}^{0}$ \cite{Sirunyan2019}. The lower bound on the $\psi$ mass is $m_{\psi}\gtrsim1.3~\text{TeV}$ which is from the gluino bound with the different group theory factor in cross section taken into account. In the case where the gluino decays to top quarks via intermediate top squark, the bound on the gluino mass is a bit stronger: $m_{\tilde{g}}\approx m_{\psi}\gtrsim1.5~\text{TeV}$ \cite{Sirunyan2019}. 

LHC searches for mass degenerate squarks bound the mass of $\phi$, since both squarks and $\phi$ decay to $j+\slashed{E}_{T}$. The recent searches at CMS place bounds on three generations of mass degenerate squarks of $m_{\tilde{q}}\gtrsim1.13~\text{TeV}$ assuming massless LSP \cite{Sirunyan2019}. Since we make the assumption of three flavors of mass degenerate exotic scalar quarks $\phi_{i}$ in WIMP cogenesis, we apply this bound directly, leading to $m_{\phi}\gtrsim1.13~\text{TeV}$.

Thus, for successful models where a matter asymmetry is produced from WIMP decays directly to baryons and ADM, the intermediate state masses are bound from below as $m_{\psi}\gtrsim m_{\phi}\sim1-2~\text{TeV}$, requiring $m_{1}\geq m_{\phi}+m_{\psi}\gtrsim3~\text{TeV}$. 

\textbf{WIMP Decay to Leptons and ADM} (Sec.~\ref{sec:lepton})

\begin{figure}[h!]
\centering
 \includegraphics[width=0.3\textwidth, trim=0 100 0 100, clip]{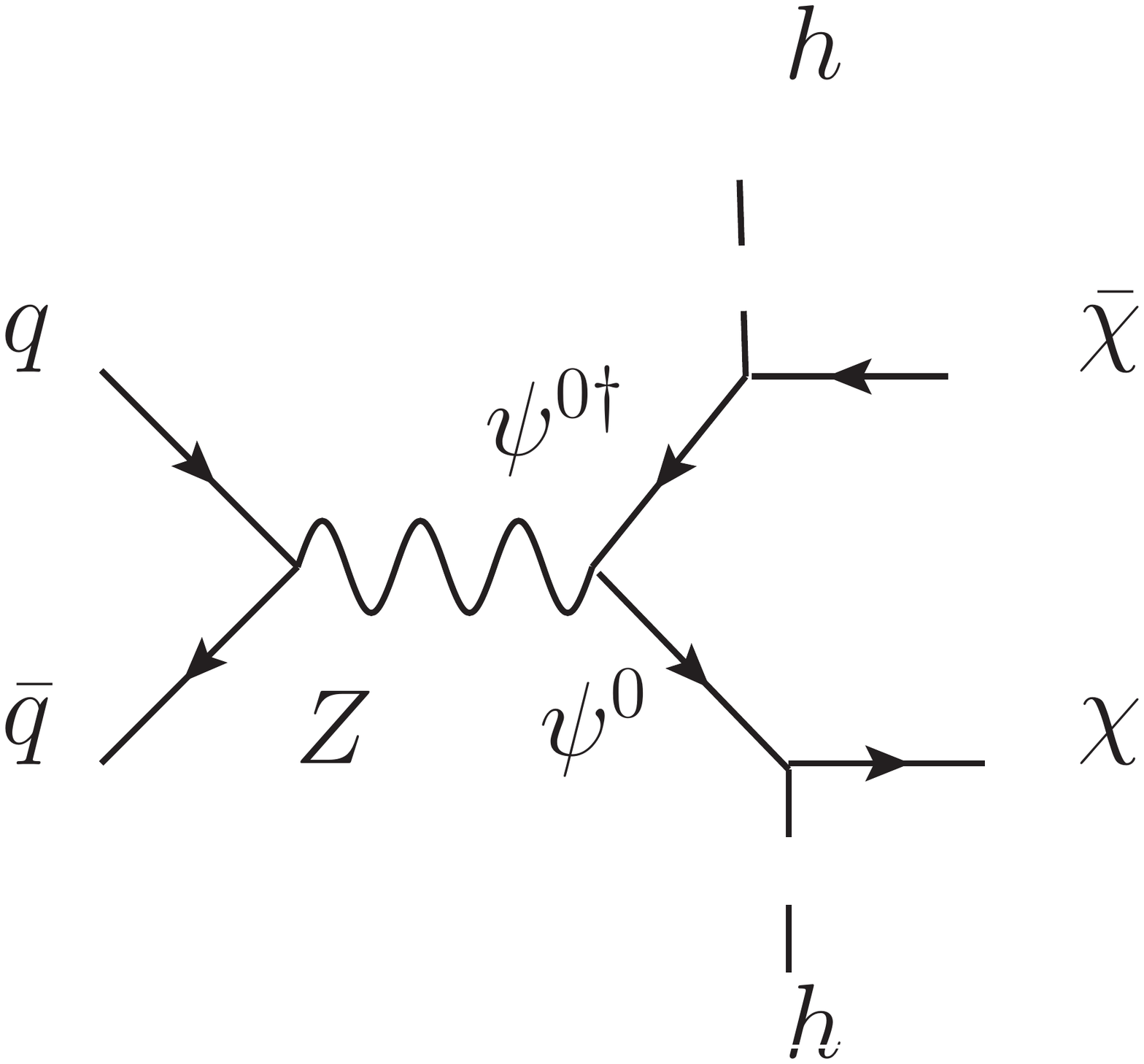}
 \includegraphics[width=0.3\textwidth, trim=0 100 0 100, clip]{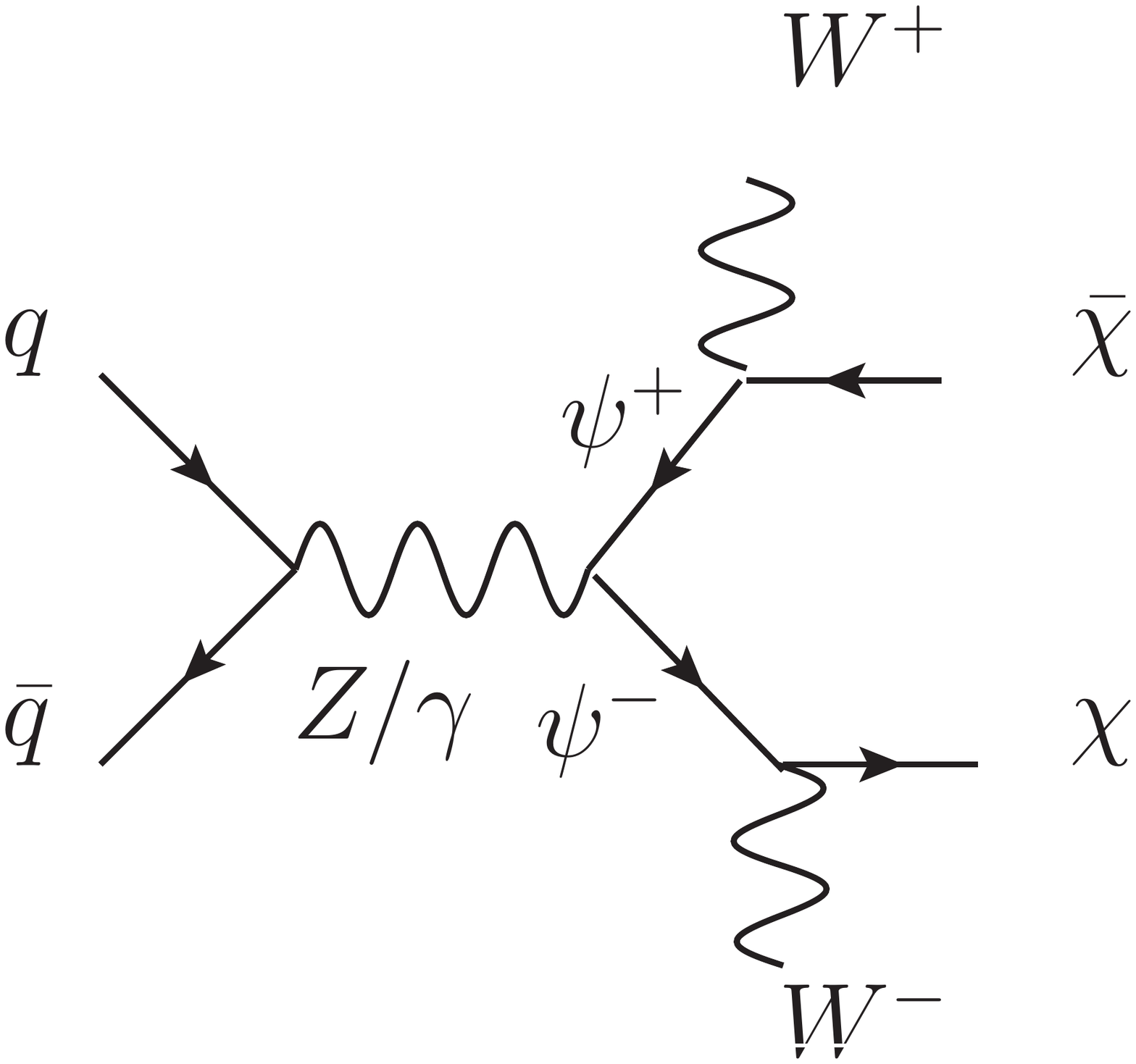}
 \includegraphics[width=0.3\textwidth, trim=0 100 0 100, clip]{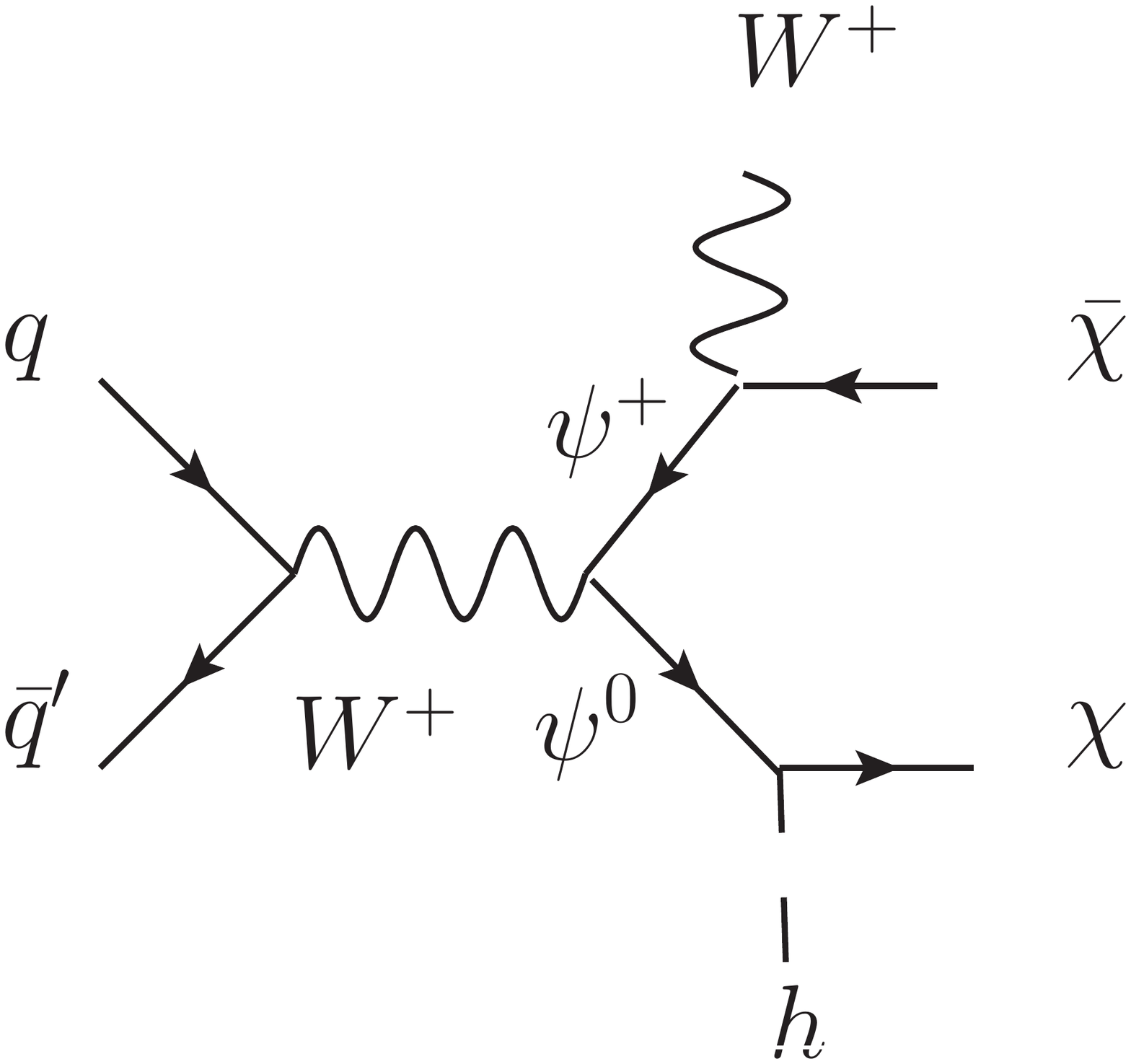}
  \caption{Diagrams relevant for $\psi$ production at hadron colliders (WIMP cogenesis with leptons).}
  \label{fig:leptoncollider}
\end{figure}
\begin{figure}[h!]
\centering
\includegraphics[width=0.3\textwidth, trim=0 130 75 130, clip]{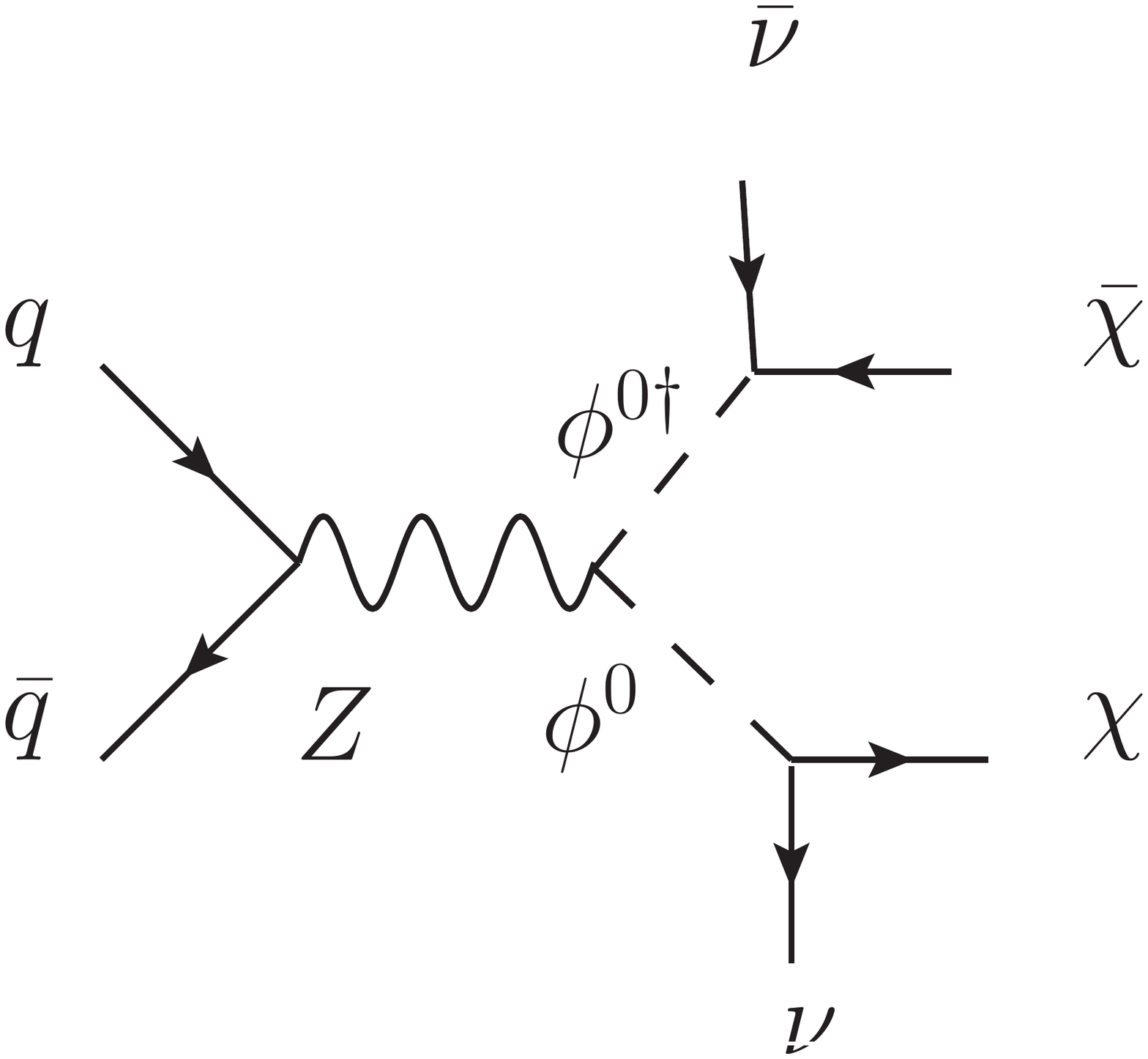}
\includegraphics[width=0.3\textwidth, trim=0 130 75 130, clip]{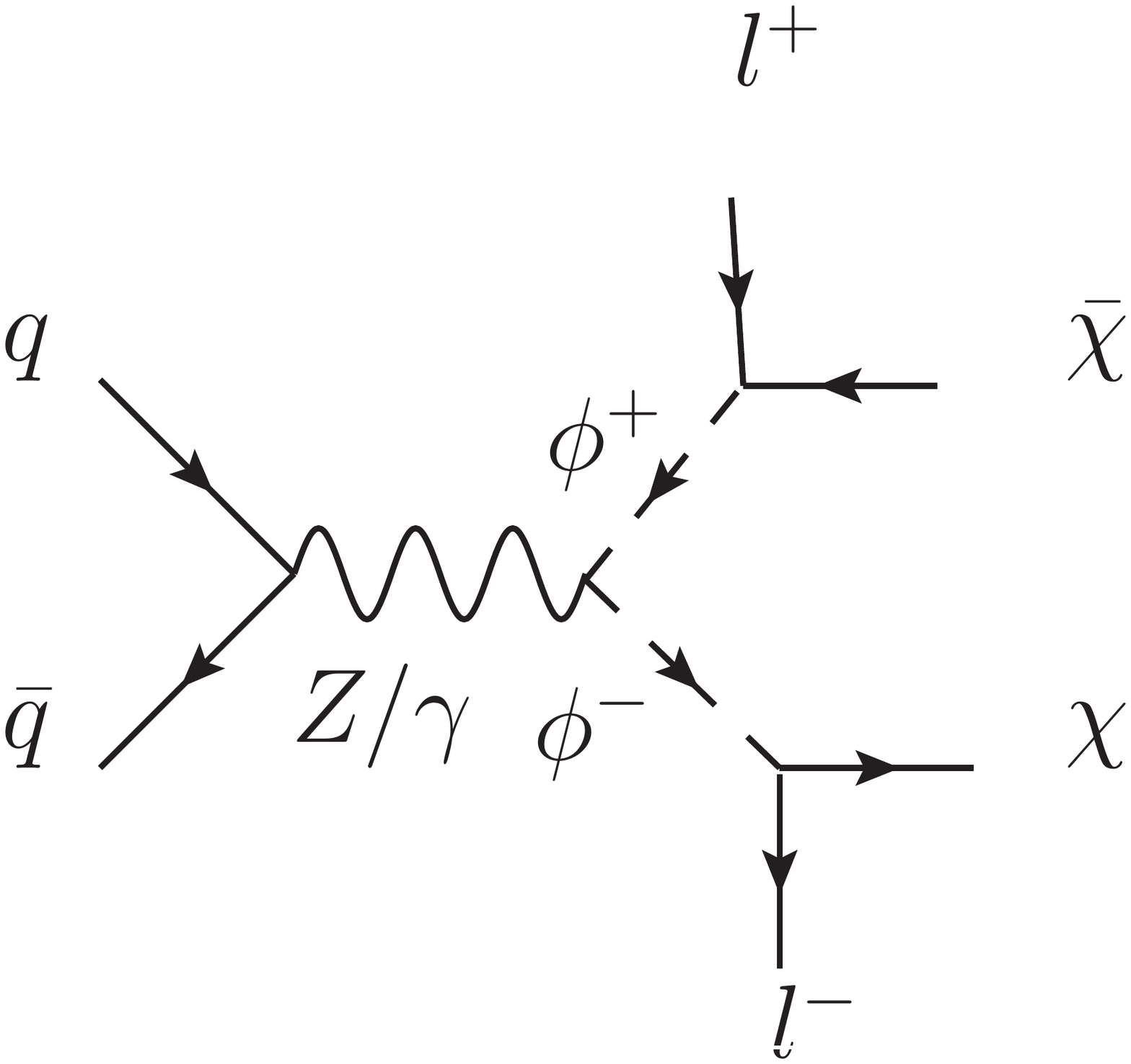}
\includegraphics[width=0.3\textwidth, trim=0 130 75 130, clip]{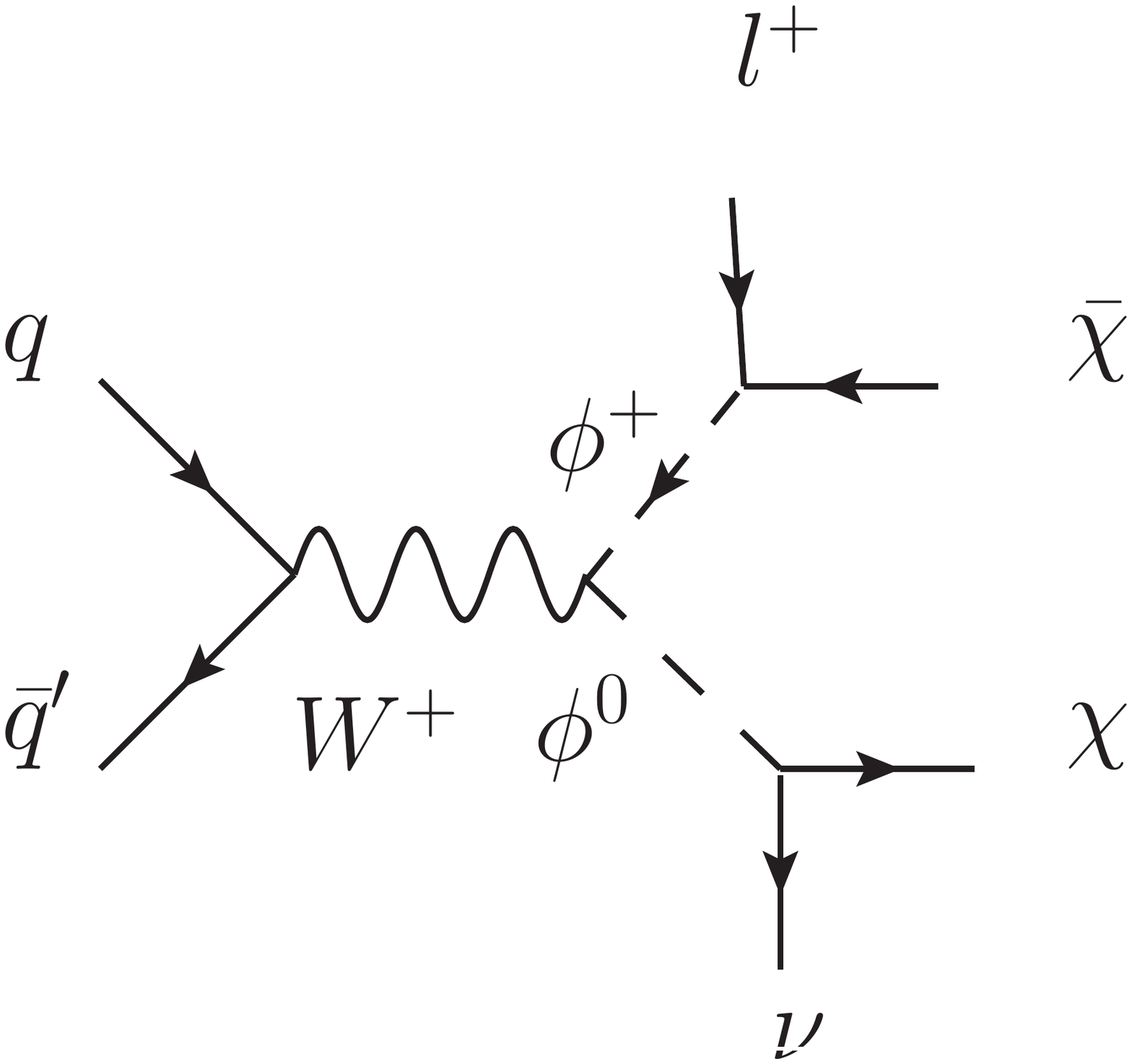}
\caption{Diagrams relevant for $\phi$ searches at hadron colliders (WIMP cogenesis with leptons).}
\label{fig:leptocollider2}
\end{figure}

In this model, $\phi$ and $\psi$ are both electroweak doublets. Thus at the LHC the neutral and charged components of these new states are produced through EW processes with intermediate $W, Z$ bosons, and subsequently decay as $\psi^0\rightarrow h\chi$, $\psi^\pm\rightarrow W^\pm\chi$, $\phi^\pm\rightarrow \ell^\pm\chi$, $\phi^0\rightarrow \nu\chi$. Consequently, these lead to signals: of $\psi^{0}\psi^{0}\rightarrow4b(4j)+\slashed{E}_{T}$, $\psi^{+}\psi^{-}\rightarrow W^{+}W^{-}+\slashed{E}_{T}$, $\phi^{+}\phi^{-}\rightarrow2\ell+\slashed{E}_{T}$, $\phi^{0}\phi^{0}\rightarrow\slashed{E}_{T}$. 
The figures for these processes are shown in Figs. \ref{fig:leptoncollider} and \ref{fig:leptocollider2}.

LHC searches for charginos $\tilde{\chi}^{\pm}$ and charged sleptons $\tilde{l}^{\pm}$ bound the charged components of $\psi,~\text{and}~\phi$, respectively, while searches for heavier neutralinos $\tilde{\chi}_{2}^{0}$ bound the neutral component of $\psi$. Specifically, searches for $\tilde{\chi}^{\pm}\rightarrow W^{\pm}\tilde{\chi}_{1}^{0}$ produces the same collider signature as decaying $\psi^{\pm}$, $\tilde{\chi}_{2}^{0}\rightarrow h\tilde{\chi}_{1}^{0}$ the same signature as decaying $\psi^{0}$, and $\tilde{l}^{\pm}\rightarrow l^{\pm}\tilde{\chi}_{1}^{0}$ the same signature as decaying $\phi^{\pm}$. Since we assume mass degeneracy among the different generations and components of $\phi$ and $\psi$, the relevant LHC searches are in the cases of $m_{\tilde{\chi}^{\pm}}=m_{\tilde{\chi}_{2}^{0}}$ and $m_{\tilde{e}}=m_{\tilde{\mu}}=m_{\tilde{\tau}}$. 

At $13~\text{TeV}$, ATLAS places bounds on the masses charginos and neutralinos with $139~\text{fb}^{-1}$ of data with $m_{\tilde{\chi}^{\pm}}=m_{\tilde{\chi}_{2}^{0}}\gtrsim740~\text{GeV}$ assuming massless LSP $\tilde{\chi}_{1}^{0}$ \cite{Aad:2019vvf}. We apply these bounds directly to the charged and neutral components of $\psi$: $m_{\psi^{\pm}}=m_{\psi^{0}}\gtrsim740~\text{GeV}$. With the same set of data ATLAS places bounds on the masses of charged sleptons in the mass degenerate limit of $m_{\tilde{l}}\gtrsim700~\text{GeV}$ \cite{Aad:2019vnb}. We apply these bounds directly to the charged components of $\phi$: $m_{\phi}\gtrsim700~\text{GeV}$.

Finally, note that just like in the earlier studied WIMP baryogenesis models \cite{wimpyBG2, Cui:2014twa}, the long-lived WIMP, $Y_1$, in WIMP cogenesis (for both the quark and lepton models we presented) is also expected to leave distinctive displaced vertex signatures if it can be produced at a collider experiment (e.g. through $qq\rightarrow Z^{'(*)}\rightarrow Y_1Y_1$). However, $Y_1$ is a SM singlet with typically O(TeV) mass which makes it hard to access with the LHC. Nevertheless it may be within reach of future high energy colliders (e.g. \cite{Benedikt:2018csr}) and leave spectacular signatures involving both displaced vertices (baryon asymmetry) and missing energy (ADM). 

A complementary signal at colliders is possible via the S mixing with the SM Higgs through $S|H|^2$ or $S^2|H|^2$ \cite{Krasnikov:1997nh}. This can lead to rare or invisible Higgs decay through $h\rightarrow SS$ \cite{Bowen:2007ia,PhysRevD.89.083513,Curtin:2013fra}. The specifics of the signal channel depends on model details about $S-H$ interactions which is beyond the scope of this work.

\subsection{Dark Matter Direct Detection}\label{sec:dd}

As expected in most of asymmetric DM models, since $\bar{\chi}$ is depleted to triviality in the early universe, indirect detection rates are negligible. Therefore we focus on the direct detection prospect of $\chi$. 

\textbf{WIMP decay to baryons and ADM (Sec.~\ref{sec:quark})}

\begin{figure}[h!]\centering
\includegraphics[width=0.4\textwidth, trim=0 250 0 250, clip]{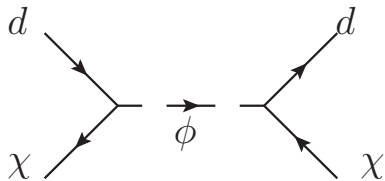}
\caption{Dominant process contributing to $\chi N\rightarrow\chi N$ scattering.}
\label{fig:quarkdd}
\end{figure}

The only available channel for $\chi$ to interact with quarks is $\chi d\rightarrow\chi d$ mediated by $\phi$. By integrating out $\phi$ in the low energy effective theory, the effective DM-quark interaction operator is 
 $\frac{\alpha_{i}^{2}}{m_{\phi}^{2}}(\bar{d}\chi)(\chi d)$, leading to spin-independent (SI) interactions between the DM and nucleon. These translate to contributions to a $\chi $-nucleon effective interaction following \cite{micromegasDD}. The SI $\chi $-nucleon cross section is 
\begin{equation}\label{eq:20}
\sigma_{\text{SI}}(\chi N\rightarrow\chi N )\approx\frac{1}{\pi}\bigg[\frac{m_{\chi }m_{n}}{m_{\phi}^{2}(m_{\chi}+m_{n})}(0.26\alpha_{s}^{2}-0.967\alpha_{d}^{2})\bigg]^{2}
\end{equation}
As we have seen, the DM mass in WIMP cogenesis model is predicted to be in the sub-GeV to GeV range. The strongest current limits on $\mathcal{O}(\text{GeV})$ SI DM-nucleon interactions come from DarkSide-50 \cite{darkside50}: for DM masses within 2-3 GeV, the upper limit on the DM-nucleon cross section is $5-7\times10^{-42}~\text{cm}^{2}$. In the case that the asymmetry is produced before the EWPT, the DM mass is below 1 GeV and the strongest bounds come from CRESST \cite{Bondarenko:2019vrb}. Specifically for DM masses of $0.5-1$ GeV, the upper limit on DM-nucleon scattering is between $\sigma_{SI}\sim10^{-38}-10^{-36}$ $\rm cm^2$.

Now we give numerical examples from our model. With $\alpha_{d}=\alpha_{s}=1$ and scalar mass at the lower bound provided by colliders, $m_{\phi}=2$ TeV and $m_{\chi}=2.5~\text{GeV}$, the SI DM-nucleon cross section is $\sigma(\chi N\rightarrow\chi N)\approx2\times10^{-42}~\text{cm}^{2}$. This is not only currently safe from the most stringent bound, but also within reach future iterations of DarkSide and other upcoming direct detection experiments \cite{Aalseth:2017fik, Essig:2017kqs, Akerib:2018dfk}. In the case that $m_{\chi}=0.89~\text{GeV}$, we again take $\alpha_{s}=\alpha_{d}=1$ and scalar masses $m_{\phi}=2~\text{TeV}$, we obtain a benchmark value from Eq. \ref{eq:20} of $\sigma_{\text{SI}}(\chi N\rightarrow\chi N)\approx1.02\times10^{-42}$ which is well below the bound set by CRESST but can be within reach of future searches for sub-GeV DM such as with the LUX-ZEPLIN\cite{Akerib:2018dfk}. 

\textbf{WIMP Decay to Leptons and ADM (Sec.~\ref{sec:lepton})}\\
In this model, the dominant process for direct detection come from tree-level $\chi -e^{-}$ scattering via $\phi$ exchange. The diagram is identical to that for ADM-nucleon scattering in the quark model, with the quarks replaced with electrons. We can estimate the cross section for ADM-electron scattering by integrating out $\phi$: 
\begin{equation}
\sigma(\chi e^{-}\rightarrow\chi e^{-})\approx\frac{1}{4\pi}\left(\frac{\alpha^{2}m_{\chi }}{m_{\phi}^{2}}\right)^{2}
\end{equation}
Similar to WIMP cogenesis with quarks, the ADM mass is fixed by the ratio of DM to baryonic matter today. In our example model of WIMP cogenesis with leptons, the ADM mass is $m_{\chi }\approx0.93-1.37~\text{GeV}$. For this mass range, Xenon100 constrains the cross-section of DM-scattering with electrons to be $\sigma_{SI}\lesssim1-2\times10^{-37}~\text{cm}^{2}$ \cite{Essig:2017kqs}. 

Owing to less stringent collider constraints, the masses of the intermediate states can be lighter in the model of WIMP cogenesis with leptons: $m_{\phi},~m_{\psi}\sim 700~\text{GeV}$. However, we need WIMP cogenesis to occur before the EWPT, when the temperature would be around or below $m_{\phi,\psi}$. Furthermore, the ADM annihilation to leptons is less constrained than annihilation to quarks \cite{admeffectiveops} and we can have $\alpha>g$. Taking the benchmark parameters of $m_{\chi}=0.93-1.37~\text{GeV}$, $m_{\phi}=700~\text{GeV}$ and $\alpha=1$ gives $\sigma(\chi e^{-}\rightarrow\chi e^{-})\approx1.1-2.4\times10^{-40}~\text{cm}^{2}$ which is just below the current bound by Xenon100 \cite{Essig:2017kqs}. 

\begin{figure}\centering
\includegraphics[width=0.5\textwidth, trim=0 240 0 240, clip]{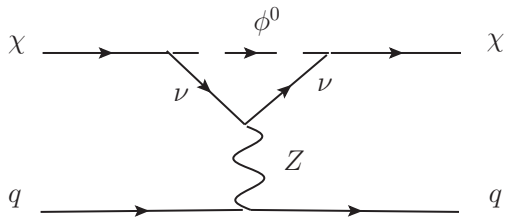}
\caption{Loop diagram contributing to direct detection rate in WIMP cogenesis with leptons. There is another diagram contributing to $\chi q\rightarrow\chi q$ with the replacements $\phi^{0}\rightarrow\phi^{\pm}~\text{and}~\nu\rightarrow l^{\pm}$. }
\label{fig:leptontriangle}
\end{figure}

There are 1-loop processes in WIMP cogenesis with leptons (Fig. \ref{fig:leptontriangle}), that allow for our sub-GeV ADM to scatter with nucleons at direct detection experiments. However, the loop suppression combined with minimal bounds on sub-GeV DM scattering with nucleons makes the rate well below the sensitivity reach of foreseeable experiments.

\subsection{Induced Nucleon Decay}\label{sec:ind}
\begin{figure}[h!]\centering
\includegraphics[width=0.5\textwidth, trim=0 250 0 250, clip]{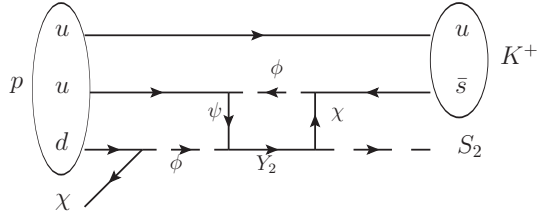}
\caption{Potential induced nucleon decay signature arising in a model of WIMP cogenesis with baryons.}
\label{fig:ind}
\end{figure}

In the model model presented in Sec.~\ref{sec:quark}, a potential signal of B-violating (induced) nucleon decay is highly suppressed and undetectable with foreseeable experiments. However, an observable induced nucleon decay (IND) signature may arise with a minimal, well-motivated extension, for instance, by introducing an additional singlet scalar, $S_2$, with $m_{S_{2}}\lesssim$ GeV. With this introduction of $S_{2}$ comes a plethora of potential interactions. Of particular interest is the Yukawa interaction $\mathcal{L}\supset \gamma S_{2}\bar{Y}_{2}P_{R}\chi$ which then requires $S_{2}$ carry $Z_{4}$ charge $i$ and generalized baryon number $G_{S_2}=1/2$. This interaction, together with the set of interactions in Eq. \ref{eq:Lyuk} allows for the possibility of induced nucleon decay, as shown in Fig. \ref{fig:ind}. The analogous diagram with $Y_1$ is much more suppressed due to the very small $\eta_1$ to ensure a long lifetime of $Y_1$.
$S_2$ can be a stable subdominant DM, or may decay, e.g. to $S$. The final decay channels from $S_2$ depend on model specifics beyond our minimal model, which we will defer for future consideration. Nevertheless a common feature is that for down-scattering processes, where $m_{\chi }>m_{S_2}$, the outgoing $K$ meson momentum from IND will be larger than those resulting from standard nucleon decays. The IND event topology here resembles that in Hylogenesis \cite{hylogenesis} while this model is fully renormalizable.

The scattering process of $p+\chi \rightarrow K^{+}+S_{2}$ effectively proceeds with a dimension-7 operator $\sim\frac{\alpha^{2}\beta\gamma \eta_{2}}{16\pi^{2}m_{2}^{3}}S_2(\bar{\chi}P_Rd)(\bar{u}P_Rd)$, and can be estimated as:
 $$\sigma(p+\chi\rightarrow S_{2}+K^{+})\sim\frac{1}{16\pi^{3}}\Big(\frac{\alpha^{2}\beta\gamma\eta_{2}m_{p}m_{\chi }}{m_{2}^{3}}\Big)^{2}$$
This leads to a prediction for the proton lifetime as $\tau_{p}^{-1}=n_{DM}\sigma(p+\chi\rightarrow S_{2}+\pi^{+})v$. This model can lead to a proton lifetime that is consistent with current lower bound set by SuperKamiokande searches \cite{superk} while within reach of future experiments such as HyperKamiokande \cite{Migenda:2017oas} and DUNE \cite{Acciarri:2016crz}. A benchmark example is: $m_{\chi}=2.5~\text{GeV}$, $m_{2}\sim3 ~\text{TeV}$, and all couplings $\sim1$ leads to proton lifetime of $\tau_{p}\sim2\times10^{36}$ years.

\subsection{Other Experimental Constraints}
As discussed in the Model Setup (Sec.~\ref{sec:model} and \ref{sec:lepton}), new sources of FCNC are absent due to the $U(3)$ flavor symmetry of the model and thus the model is consistent with related constraints on FCNC. In addition, despite the presence of CP violation source necessary for the asymmetry generation, the model is exempt from the constraints on electric dipole moments (EDMs) for the neutron and electron \cite{Abel:2020gbr, Andreev:2018ayy} . The reason is that, the interference diagrams (Fig.~\ref{fig:cpdiagrams}) leading to CP violation do not involve SM quarks or leptons, and the new fields couple exclusively to right-handed quarks or left-handed leptons. 

WIMP cogenesis with baryons evades bounds from neutron-antineutron oscillation: the intrinsic interactions in the model and the $U(3)$ flavor symmetry together forbid $udd\rightarrow\bar{u}\bar{d}\bar{d}$ conversion at tree-level and 1-loop (alternatively with small couplings without invoking the flavor symmetry). Higher order process is strongly suppressed by loop factors and the TeV-scale masses of $Y_{1,2}$, $\psi$, and $\phi$, even with $\mathcal{O}(1)$ couplings. 

\section{Conclusion}\label{sec:fin}
In this paper we proposed WIMP cogenesis, a novel mechanism which addresses the \textit{tripple puzzle} about cosmic matter abundance in a unified framework: asymmetric dark matter and a baryon or lepton asymmetry are simultaneously generated from the same decay chain of a freezeout population of metastable WIMPs. The WIMP plays the role of grandparent for the matter abundance in the Universe, meanwhile the ``coincidence" between DM and baryon abundances is automatically addressed via their co-production. Additionally, the WIMP decay chain readily permits DM and baryon asymmetries to inherit a generalized WIMP miracle. The three Sakharov conditions are satisfied in three subsequent stages in order. ADM and baryons (leptons) share a generalized baryon (lepton) number symmetry that is conserved. We present two renormalizable models as benchmark examples realizing the idea, and find that with perturbative couplings and weak-scale masses for the new states, the observed DM and baryon relic densities can be explained while being compatible with relevant constraints. The models neatly predict ADM with mass $m_{DM}\sim0.7-2.5~\text{GeV}$. These models can lead to testable signatures at a variety of experiments, including (low mass) DM direct detection, nucleon decay and the production of new SM charged particles at the LHC. Furthermore the long-lived WIMP in these models may be accessible with future high energy colliders, leaving spectacular signals by reproducing the cogenesis of matter in the early Universe.

\section*{Acknowledgements}

We thank Matthew Dolan, Aniket Joglekar and Brian Shuve for discussions. We thank Brian Shuve and Raman Sundrum for commenting on the manuscript. Feynman diagrams were drawn using JaxoDraw \cite{Binosi:2008ig}. The authors are supported in part by the US Department of Energy grant DE-SC0008541. YC thanks the Kavli Institute for Theoretical Physics (supported by the National Science Foundation under Grant No. NSF PHY-1748958) for the support and hospitality while the work was being completed.

 \appendix
\section{Relating Baryon and Lepton Asymmetries for WIMP Cogenesis before Electroweak Phase Transition}\label{sec:cs}
In this Appendix we derive the relation between baryon and lepton asymmetries for WIMP cogenesis before electroweak phase transition. We will follow the general procedure laid out for the SM \cite{Riotto:1998bt, Buchmuller:2005eh, tasi}, while adding in the effects from new particles in WIMP cogenesis models.

\subsection{WIMP Decay to Baryons and ADM (Sec. \ref{sec:quark})}\label{sec:cs1}
Before the electroweak phase transition (EWPT), chemical equilibrium of SM left-handed and right-handed quarks and leptons, Higgs bosons, and new fields introduced by WIMP cogenesis $\phi$, $\psi$, and $\chi$ determines the relationship between number densities of baryons, leptons, and ADM candidate $\chi$. This relationship and the observed ratio $\Omega_{\chi}/\Omega_{B}\approx5$ determines the ADM mass as in Eq. \ref{eq:admmass}. In the high temperature plasma of the early universe the quarks, leptons, Higgs, $\phi$, $\psi$, and $\chi$ interact via gauge, Yukawa, and sphaleron processes. The interactions that constrain the chemical potentials in thermal equilibrium are:
\begin{enumerate}
\item The effective sphaleron interaction $\mathcal{O}_{\text{sph}}\sim\displaystyle\prod_{i}(Q_{i}Q_{i}Q_{i}L_{i})$ gives rise to 
\begin{equation}\label{eq:sphaleron}
\sum_{i}(3\mu_{Q_{i}}+\mu_{L_{i}})=0
\end{equation}
where $i$ is an index counting the number of generations of fermions and $Q_{i}$ are the LH quarks and $L_{i}$ are the LH leptons.
\item The $SU(3)$ QCD instanton processes lead to interactions between LH quarks and RH quarks $u_{i}$ and $d_{i}$. These interactions are described by $\mathcal{O}_{\text{inst}}\sim\displaystyle\prod_{i}(Q_{i}Q_{i}u_{i}^{c}d_{i}^{c})$ which leads to
\begin{equation}\label{eq:instanton}
\sum_{i}(2\mu_{Q_{i}}-\mu_{u_{i}}-\mu_{d_{i}})=0
\end{equation}
\item The total hypercharge of the plasma must vanish at all temperatures. In addition to the hypercharge carried by SM states, $\phi$ and $\psi$ also contribute, while the magnitude of the contribution depends on their masses relative to EWPT temperature $T_{\rm EWPT}$. Non-relativistic $\phi$ and $\psi$ bear a Boltzmann suppression in their equilibrium density distribution which makes their contribution to hypercharge density negligible relative to relativistic species. Given the unknowns around determining $T_{\rm EWPT}$ and the wide ranges $m_\psi\sim m_\phi$, we consider possibilities at two limits: $m_\psi\sim m_\phi\ll T_{\rm EWPT}$ and $m_\psi\sim m_\phi\gg T_{\rm EWPT}$. With $m_\psi\sim m_\phi\ll T_{\rm EWPT}$, we have:
\begin{equation}\label{eq:hypercharge1}
\sum_{i}(\mu_{Q_{i}}+2\mu_{u_{i}}-\mu_{d_{i}}-\mu_{L_{i}}-\mu_{e_{i}}+\frac{2N_{H}}{N_{f}}\mu_{H}+\mu_{\phi_{i}}+\mu_{\psi_{i}})=0
\end{equation}
where $N_{H}$ is the number of Higgs bosons (1 in the SM) and $N_{f}$ is the number of generations of fermions.
With $m_\psi\sim m_\phi\gg T_{\rm EWPT}$, we have:
\begin{equation}\label{eq:Ycharge2}
\sum_{i}(\mu_{Q_{i}}+2\mu_{u_{i}}-\mu_{d_{i}}-\mu_{L_{i}}-\mu_{e_{i}}+\frac{2N_{H}}{N_{f}}\mu_{H})=0
\end{equation}

\item The Yukawa interactions of the SM $\mathcal{O}_{\text{SM}}\sim\bar{Q}_{i}Hd_{j},~\bar{Q}_{i}\tilde{H}u_{j},~\bar{L}_{i}He_{j}$ and the Yukawa interactions introduced in Sec \ref{sec:model} $\mathcal{O}_{\text{WIMP}}\sim\phi_{i}\bar{d}_{i}\chi^{c},~\beta_{ijk}\phi_{i}\bar{\psi}_{j}u_{k}$, while in equilibrium give rise to 
\begin{equation}\label{eq:yukawa}
\begin{aligned}
\mu_{Q_{i}}-\mu_{H}-\mu_{d_{j}}=0\\ 
\mu_{Q_{i}}+\mu_{H}-\mu_{u_{j}}=0\\ 
\mu_{L_{i}}-\mu_{H}-\mu_{e_{j}}=0\\ 
\mu_{d_{i}}-\mu_{\phi_{i}}+\mu_{\chi}=0\\ 
\mu_{\psi_{j}}-\mu_{\phi_{i}}-\mu_{u_{k}}=0
\end{aligned}
\end{equation}
\end{enumerate}
Since the temperature before the EWPT is much greater than the masses of the quarks, leptons, and $\chi$ we take the massless limit where their number densities are $n_{i}-\bar{n}_{i}=\frac{1}{6}g\mu_{i}T^{2}$. The baryon, lepton, and $\chi$ number densities are $n_{B}=\frac{1}{6}BT^{2}$, $n_{L}=\frac{1}{6}LT^{2}$, and $n_{X}=\frac{1}{6}XT^{2}$, respectively, where
\begin{eqnarray}\label{eq:blx}
B&=&\sum_{i}(2\mu_{Q_{i}}+\mu_{u_{i}}+\mu_{d_{i}})\\
L&=&\sum_{i}(2\mu_{L_{i}}+\mu_{e_{i}})\\
X&=&\mu_{\chi}
\end{eqnarray}
With SM alone, the combination of asymmetry $B-L$ is preserved, while in our model $B-L+2X$ would be preserved. Assuming equilibrium amongst the various generations $\mu_{Q_{i}}\equiv\mu_{Q}$, $\mu_{L_{i}}\equiv\mu_{L}$, $\mu_{e_i}\equiv\mu_e$, $\mu_{q_i}\equiv\mu_{q}$, $\mu_{\phi_i}\equiv\mu_{\phi}$, $\mu_{\psi_i}\equiv\mu_{\psi}$ allows us to write $B=N_{f}(2\mu_{Q}+\mu_{u}+\mu_{d})$, $L=N_{f}(2\mu_{L}+\mu_{e})$. Thus the preserved combination, per generation, is 
\begin{equation}\label{eq:blxcons}
\big[2\mu_{Q}+\mu_{u}+\mu_{d}-(2\mu_{L}+\mu_{e})\big]+2\mu_{\chi}=0
\end{equation}

Let us first analyze the case of $m_{\psi}\sim m_{\phi}\ll T_{\text{EWPT}}$. Using the Yukawa interactions of Eqs. \ref{eq:yukawa} , Eq. \ref{eq:blxcons} can be recast as $\mu_{\chi}=-\frac{1}{2}(B-L)=-\frac{1}{2}(13\mu_{Q}+\mu_{H})=\mu_{\phi}-\mu_{Q}+\mu_{H}$. The effective sphaleron interactions of Eq. \ref{eq:sphaleron} give $\mu_{L}=-3\mu_{Q}$. Substituting this and Eqs. \ref{eq:yukawa} in Eq. \ref{eq:hypercharge1} allows us to solve $\mu_{H}$ in terms of $\mu_Q$ which allows us to write all chemical potentials in terms of $\mu_{Q}$ using Eqs. \ref{eq:yukawa}:
\begin{equation}\label{eq:bgchem}
\begin{aligned}
&\mu_{L}=-3\mu_{Q} &&\mu_{H}=\frac{N_{f}}{N_{f}+N_{H}}\mu_{Q}\\ &\mu_{u}=\frac{2N{f}+N_{H}}{N_{f}+N_{H}}\mu_{Q} &&\mu_{d}=\frac{N_{H}}{N_{f}+N_{H}}\mu_{Q}\\ &\mu_{e}=-\frac{4N_{f}+3N_{H}}{N_{f}+N_{H}}\mu_{Q} &&\mu_{\phi}=-\frac{1}{2}\bigg(\frac{14N_{f}+11N_{H}}{N_{f}+N_{H}}\bigg) \mu_{Q}\\ &\mu_{\psi}=-\frac{1}{2}\bigg(\frac{10N_{f}+9N_{H}}{N_{f}+N_{H}}\bigg) \mu_{Q} &&\mu_{\chi}=-\frac{1}{2}\bigg(\frac{14N_{f}+13N_{H}}{N_{f}+N_{H}}\bigg) \mu_{Q}
\end{aligned}
\end{equation}

Plugging these into the equations for $B$, $L$ and $B-L$ allows us to write the relations between them:
\begin{eqnarray}\label{eq:bgrelation}
B&=&4N_{f}\mu_{Q}\\ L&=&-\frac{10N_{f}+9N_{H}}{N_{f}+N_{H}}N_{f}\mu_{Q}\\ B-L&=&\frac{14N_{f}+13N_{H}}{N_{f}+N_{H}}N_{f}\mu_{Q}\equiv c_{s}^{-1}B
\end{eqnarray}
where 
\begin{equation}\label{eq:bgcs}
c_{s}\equiv B/(B-L)=\frac{4(N_{f}+N_{H})}{14N_{f}+13N_{H}}
\end{equation}

In the other limit, $m_{\psi}\sim m_{\phi}\gg T_{\text{EWPT}}$, we use Eq. \ref{eq:Ycharge2}. In this case, we need only use the SM Yukawa interactions to find the SM chemical potentials (and thus $c_{s}\equiv B/B-L$). We can still use Eq. \ref{eq:blxcons} to find the chemical potentials of $\phi,~\psi,~\text{and}~\chi$ in terms of $\mu_Q$: 
\begin{equation}\label{eq:bgchemheavy}
\begin{aligned}
&\mu_{L}=-3\mu_{Q} &&\mu_{H}=-\frac{4N_{f}}{2N_{f}+N_{H}}\mu_{Q}\\ &\mu_{u}=-\frac{2N_{f}-N_{H}}{2N_{f}+N_{H}}\mu_{Q} &&\mu_{d}=\frac{6N_{f}+N_{H}}{2N_{f}+N_{H}}\mu_{Q}\\ &\mu_{e}=-\frac{2N_{f}+3N_{H}}{2N_{f}+N_{H}}\mu_{Q} &&\mu_{\phi}=-\frac{1}{2}\bigg(\frac{10N_{f}+11N_{H}}{2N_{f}+N_{H}}\bigg) \mu_{Q}\\ &\mu_{\psi}=-\frac{1}{2}\bigg(\frac{14N_{f}+9N_{H}}{2N_{f}+N_{H}}\bigg) \mu_{Q} &&\mu_{\chi}=-\frac{1}{2}\bigg(\frac{22N_{f}+13N_{H}}{2N_{f}+N_{H}}\bigg) \mu_{Q}
\end{aligned}
\end{equation}
Plugging these into the same equations for $B,~L,~\text{and}~B-L$ yields 
\begin{eqnarray}\label{eq:bgrelationheavy}
B&=4&N_{f}\mu_{Q}\\ L&=&-\frac{14N_{f}+9N_{H}}{2N_{f}+N_{H}}N_{f}\mu_{Q}\\ B-L&=&\frac{22N_{f}+13N_{H}}{2N_{f}+N_{H}}N_{f}\mu_{Q}\equiv c_{s}^{-1}B
\end{eqnarray}
where 
\begin{equation}\label{eq:bgcsheavy}
c_{s}\equiv B/(B-L)=\frac{8N_{f}+4N_{H}}{22N_{f}+13N_{H}}
\end{equation}
which is the same as the result in the SM \cite{tasi}. 

\subsection{WIMP Decay to Leptons and ADM (Sec. \ref{sec:lepton})} \label{sec:cs2}
In the model outlined in Sec.~\ref{sec:lepton} WIMP cogenesis, the biggest change is to the Yukawa interactions: $\mathcal{O}_{\text{WIMP}}\sim\phi\bar{L}\chi^{c},~H\bar{\psi}\chi$ which changes the last two Yukawa interactions in Eqs. \ref{eq:yukawa} in a straightforward fashion. We note also the mass of the ADM candidate $\chi$ is fixed by the observed ratio of DM to baryon energy densities fixed by: \begin{multline*}\Omega_{DM}=\frac{2m_{\chi}s_{0}}{\rho_{0}}\epsilon_{1}Y_{Y_{1},f.o.}=5\Omega_{B}=\frac{5c_{s}s_{0}m_{n}}{|c_{s}-1|\rho_{0}}\epsilon_{1}Y_{Y_{1},f.o.}\implies m_{\chi}=\frac{5c_{s}}{2|c_{s}-1|}m_{n}\end{multline*} Following the same procedure, we find the chemical potentials in terms of $\mu_Q$ to be  
\begin{equation}\label{eq:lgchem}
\begin{aligned}
&\mu_{L}=-3\mu_{Q} &&\mu_{H}=\frac{4N_{f}}{2N_{f}+N_{H}}\mu_{Q}\\ &\mu_{d}=-\frac{2N_{f}-N_{H}}{2N_{f}+N_{H}}\mu_{Q} &&\mu_{u}=\frac{6N_{f}+N_{H}}{2N_{f}+N_{H}}\mu_{Q}\\ &\mu_{e}=-\frac{10N_{f}+3N_{H}}{2N_{f}+N_{H}}\mu_{Q} &&\mu_{\phi}=-\frac{1}{2}\bigg(\frac{42N_{f}+19N_{H}}{2N_{f}+N_{H}}\bigg) \mu_{Q}\\ &\mu_{\chi}=-\frac{1}{2}\bigg(\frac{30N_{f}+13N_{H}}{2N_{f}+N_{H}}\bigg) \mu_{Q} &&\mu_{\psi}=-\frac{1}{2}\bigg(\frac{22N_{f}+13N_{H}}{2N_{f}+N_{H}}\bigg) \mu_{Q}
\end{aligned}
\end{equation}

Again, following the same procedure as before we find 

\begin{eqnarray}\label{lgrelation}
B&=&4N_{f}\mu_{Q}\\ L&=&-\frac{22N_{f}+9N_{H}}{2N_{f}+N_{H}}N_{f}\mu_{Q}\\ B-L&=&\frac{30N_{f}+13N_{H}}{2N_{f}+N_{H}}N_{f}\mu_{Q}\equiv c_{s}^{-1}B
\end{eqnarray}
where
\begin{equation}\label{eq:cslg}
c_{s}\equiv\frac{B}{B-L}=\frac{8N_{f}+4N_{H}}{30N_{f}+13N_{H}}
\end{equation}

In the case that $\phi$ and $\psi$ are heavy, the same result as that given in Eq. \ref{eq:bgcsheavy} is found, but the chemical potentials of $\phi$, $\psi$ and $\chi$ are

\begin{eqnarray*}
\mu_{\phi}=-\frac{1}{2}\bigg(\frac{34N_{f}+19N_{H}}{2N_{f}+N_{H}}\bigg)\\
\mu_{\psi}=-\frac{1}{2}\bigg(\frac{30N_{f}+13N_{H}}{2N_{f}+N_{H}}\bigg)\\
\mu_{\chi}=-\frac{1}{2}\bigg(\frac{22N_{f}+13N_{H}}{2N_{f}+N_{H}}\bigg).
\end{eqnarray*}

\bibliographystyle{JHEP}
\bibliography{WIMPcogenesis_refs}

\providecommand{\noopsort}[1]{}\providecommand{\singleletter}[1]{#1}%

\providecommand{\href}[2]{#2}\begingroup\raggedright\begin{thebibliography}{10}

\bibitem{planck}
{\scshape Planck} collaboration, \emph{{Planck 2018 results. VI. Cosmological
  parameters}},  \href{https://arxiv.org/abs/1807.06209}{{\ttfamily
  1807.06209}}.

\bibitem{darkside50}
{\scshape DarkSide} collaboration, \emph{{Low-Mass Dark Matter Search with the
  DarkSide-50 Experiment}},
  \href{https://doi.org/10.1103/PhysRevLett.121.081307}{\emph{Phys. Rev. Lett.}
  {\bfseries 121} (2018) 081307}
  [\href{https://arxiv.org/abs/1802.06994}{{\ttfamily 1802.06994}}].

\bibitem{indirect}
G.~Bertone, N.~Bozorgnia, J.~S. Kim, S.~Liem, C.~McCabe, S.~Otten et~al.,
  \emph{{Identifying WIMP dark matter from particle and astroparticle data}},
  \href{https://doi.org/10.1088/1475-7516/2018/03/026}{\emph{JCAP} {\bfseries
  1803} (2018) 026} [\href{https://arxiv.org/abs/1712.04793}{{\ttfamily
  1712.04793}}].

\bibitem{Mitsou:2014wta}
V.~A. Mitsou, \emph{{Overview of searches for dark matter at the LHC}},
  \href{https://doi.org/10.1088/1742-6596/651/1/012023}{\emph{J. Phys. Conf.
  Ser.} {\bfseries 651} (2015) 012023}
  [\href{https://arxiv.org/abs/1402.3673}{{\ttfamily 1402.3673}}].

\bibitem{Nussinov:1985xr}
S.~Nussinov, \emph{{TECHNOCOSMOLOGY: COULD A TECHNIBARYON EXCESS PROVIDE A
  'NATURAL' MISSING MASS CANDIDATE?}},
  \href{https://doi.org/10.1016/0370-2693(85)90689-6}{\emph{Phys. Lett.}
  {\bfseries 165B} (1985) 55}.

\bibitem{Barr:1990ca}
S.~M. Barr, R.~S. Chivukula and E.~Farhi, \emph{{Electroweak Fermion Number
  Violation and the Production of Stable Particles in the Early Universe}},
  \href{https://doi.org/10.1016/0370-2693(90)91661-T}{\emph{Phys. Lett.}
  {\bfseries B241} (1990) 387}.

\bibitem{Kaplan:1991ah}
D.~B. Kaplan, \emph{{A Single explanation for both the baryon and dark matter
  densities}}, \href{https://doi.org/10.1103/PhysRevLett.68.741}{\emph{Phys.
  Rev. Lett.} {\bfseries 68} (1992) 741}.

\bibitem{Kaplan:2009ag}
D.~E. Kaplan, M.~A. Luty and K.~M. Zurek, \emph{{Asymmetric Dark Matter}},
  \href{https://doi.org/10.1103/PhysRevD.79.115016}{\emph{Phys. Rev.}
  {\bfseries D79} (2009) 115016}
  [\href{https://arxiv.org/abs/0901.4117}{{\ttfamily 0901.4117}}].

\bibitem{Zurek:2013wia}
K.~M. Zurek, \emph{{Asymmetric Dark Matter: Theories, Signatures, and
  Constraints}},
  \href{https://doi.org/10.1016/j.physrep.2013.12.001}{\emph{Phys. Rept.}
  {\bfseries 537} (2014) 91} [\href{https://arxiv.org/abs/1308.0338}{{\ttfamily
  1308.0338}}].

\bibitem{adm}
K.~Petraki and R.~R. Volkas, \emph{{Review of asymmetric dark matter}},
  \href{https://doi.org/10.1142/S0217751X13300287}{\emph{Int. J. Mod. Phys.}
  {\bfseries A28} (2013) 1330028}
  [\href{https://arxiv.org/abs/1305.4939}{{\ttfamily 1305.4939}}].

\bibitem{wimpyBG}
Y.~Cui, L.~Randall and B.~Shuve, \emph{{A WIMPy Baryogenesis Miracle}},
  \href{https://doi.org/10.1007/JHEP04(2012)075}{\emph{JHEP} {\bfseries 04}
  (2012) 075} [\href{https://arxiv.org/abs/1112.2704}{{\ttfamily 1112.2704}}].

\bibitem{McDonald:2011zza}
J.~McDonald, \emph{{Baryomorphosis: Relating the Baryon Asymmetry to the 'WIMP
  Miracle'}}, \href{https://doi.org/10.1103/PhysRevD.83.083509}{\emph{Phys.
  Rev.} {\bfseries D83} (2011) 083509}
  [\href{https://arxiv.org/abs/1009.3227}{{\ttfamily 1009.3227}}].

\bibitem{Davidson:2012fn}
S.~Davidson and M.~Elmer, \emph{{Similar Dark Matter and Baryon abundances with
  TeV-scale Leptogenesis}},
  \href{https://doi.org/10.1007/JHEP10(2012)148}{\emph{JHEP} {\bfseries 10}
  (2012) 148} [\href{https://arxiv.org/abs/1208.0551}{{\ttfamily 1208.0551}}].

\bibitem{wimpyBG2}
Y.~Cui and R.~Sundrum, \emph{{Baryogenesis for weakly interacting massive
  particles}}, \href{https://doi.org/10.1103/PhysRevD.87.116013}{\emph{Phys.
  Rev.} {\bfseries D87} (2013) 116013}
  [\href{https://arxiv.org/abs/1212.2973}{{\ttfamily 1212.2973}}].

\bibitem{Cui:2015eba}
Y.~Cui, \emph{{A Review of WIMP Baryogenesis Mechanisms}},
  \href{https://doi.org/10.1142/S0217732315300281}{\emph{Mod. Phys. Lett.}
  {\bfseries A30} (2015) 1530028}
  [\href{https://arxiv.org/abs/1510.04298}{{\ttfamily 1510.04298}}].

\bibitem{Farina:2016ndq}
M.~Farina, A.~Monteux and C.~S. Shin, \emph{{Twin mechanism for baryon and dark
  matter asymmetries}},
  \href{https://doi.org/10.1103/PhysRevD.94.035017}{\emph{Phys. Rev.}
  {\bfseries D94} (2016) 035017}
  [\href{https://arxiv.org/abs/1604.08211}{{\ttfamily 1604.08211}}].

\bibitem{Racker:2014uga}
J.~Racker and N.~Rius, \emph{{Helicitogenesis: WIMPy baryogenesis with sterile
  neutrinos and other realizations}},
  \href{https://doi.org/10.1007/JHEP11(2014)163}{\emph{JHEP} {\bfseries 11}
  (2014) 163} [\href{https://arxiv.org/abs/1406.6105}{{\ttfamily 1406.6105}}].

\bibitem{Cui:2013bta}
Y.~Cui, \emph{{Natural Baryogenesis from Unnatural Supersymmetry}},
  \href{https://doi.org/10.1007/JHEP12(2013)067}{\emph{JHEP} {\bfseries 12}
  (2013) 067} [\href{https://arxiv.org/abs/1309.2952}{{\ttfamily 1309.2952}}].

\bibitem{Cui:2014twa}
Y.~Cui and B.~Shuve, \emph{{Probing Baryogenesis with Displaced Vertices at the
  LHC}}, \href{https://doi.org/10.1007/JHEP02(2015)049}{\emph{JHEP} {\bfseries
  02} (2015) 049} [\href{https://arxiv.org/abs/1409.6729}{{\ttfamily
  1409.6729}}].

\bibitem{Cui:2016rqt}
Y.~Cui, T.~Okui and A.~Yunesi, \emph{{LHC Signatures of WIMP-triggered
  Baryogenesis}}, \href{https://doi.org/10.1103/PhysRevD.94.115022}{\emph{Phys.
  Rev.} {\bfseries D94} (2016) 115022}
  [\href{https://arxiv.org/abs/1605.08736}{{\ttfamily 1605.08736}}].

\bibitem{ATLAS:2019ems}
{\scshape ATLAS} collaboration, \emph{{Search for long-lived, massive particles
  in events with a displaced vertex and a displaced muon in $pp$ collisions at
  $\sqrt{s} = 13$ TeV with the ATLAS detector}}, .

\bibitem{dmfrombaryonasymmetry}
R.~Kitano and I.~Low, \emph{Dark matter from baryon asymmetry},
  \href{https://doi.org/10.1103/PhysRevD.71.023510}{\emph{Phys. Rev. D}
  {\bfseries 71} (2005) 023510}.

\bibitem{admfromlepto}
A.~Falkowski, J.~T. Ruderman and T.~Volansky, \emph{{Asymmetric Dark Matter
  from Leptogenesis}},
  \href{https://doi.org/10.1007/JHEP05(2011)106}{\emph{JHEP} {\bfseries 05}
  (2011) 106} [\href{https://arxiv.org/abs/1101.4936}{{\ttfamily 1101.4936}}].

\bibitem{hylogenesis}
H.~Davoudiasl, D.~E. Morrissey, K.~Sigurdson and S.~Tulin, \emph{{Hylogenesis:
  A Unified Origin for Baryonic Visible Matter and Antibaryonic Dark Matter}},
  \href{https://doi.org/10.1103/PhysRevLett.105.211304}{\emph{Phys. Rev. Lett.}
  {\bfseries 105} (2010) 211304}
  [\href{https://arxiv.org/abs/1008.2399}{{\ttfamily 1008.2399}}].

\bibitem{Agashe:2004bm}
K.~Agashe and G.~Servant, \emph{{Baryon number in warped GUTs: Model building
  and (dark matter related) phenomenology}},
  \href{https://doi.org/10.1088/1475-7516/2005/02/002}{\emph{JCAP} {\bfseries
  0502} (2005) 002} [\href{https://arxiv.org/abs/hep-ph/0411254}{{\ttfamily
  hep-ph/0411254}}].

\bibitem{Cui:2011qe}
Y.~Cui, L.~Randall and B.~Shuve, \emph{{Emergent Dark Matter, Baryon, and
  Lepton Numbers}}, \href{https://doi.org/10.1007/JHEP08(2011)073}{\emph{JHEP}
  {\bfseries 08} (2011) 073} [\href{https://arxiv.org/abs/1106.4834}{{\ttfamily
  1106.4834}}].

\bibitem{Fonseca:2015rwa}
N.~Fonseca, L.~Necib and J.~Thaler, \emph{{Dark Matter, Shared Asymmetries, and
  Galactic Gamma Ray Signals}},
  \href{https://doi.org/10.1088/1475-7516/2016/02/052}{\emph{JCAP} {\bfseries
  1602} (2016) 052} [\href{https://arxiv.org/abs/1507.08295}{{\ttfamily
  1507.08295}}].

\bibitem{Elor:2018twp}
G.~Elor, M.~Escudero and A.~Nelson, \emph{{Baryogenesis and Dark Matter from
  $B$ Mesons}}, \href{https://doi.org/10.1103/PhysRevD.99.035031}{\emph{Phys.
  Rev.} {\bfseries D99} (2019) 035031}
  [\href{https://arxiv.org/abs/1810.00880}{{\ttfamily 1810.00880}}].

\bibitem{Kahlhoefer:2015bea}
F.~Kahlhoefer, K.~Schmidt-Hoberg, T.~Schwetz and S.~Vogl, \emph{{Implications
  of unitarity and gauge invariance for simplified dark matter models}},
  \href{https://doi.org/10.1007/JHEP02(2016)016}{\emph{JHEP} {\bfseries 02}
  (2016) 016} [\href{https://arxiv.org/abs/1510.02110}{{\ttfamily
  1510.02110}}].

\bibitem{Cui:2017juz}
Y.~Cui and F.~D'Eramo, \emph{{Surprises from complete vector portal theories:
  New insights into the dark sector and its interplay with Higgs physics}},
  \href{https://doi.org/10.1103/PhysRevD.96.095006}{\emph{Phys. Rev. D}
  {\bfseries 96} (2017) 095006}
  [\href{https://arxiv.org/abs/1705.03897}{{\ttfamily 1705.03897}}].

\bibitem{Graesser:2011wi}
M.~L. Graesser, I.~M. Shoemaker and L.~Vecchi, \emph{{Asymmetric WIMP dark
  matter}}, \href{https://doi.org/10.1007/JHEP10(2011)110}{\emph{JHEP}
  {\bfseries 10} (2011) 110} [\href{https://arxiv.org/abs/1103.2771}{{\ttfamily
  1103.2771}}].

\bibitem{tasi}
M.-C. Chen, \emph{{TASI 2006 Lectures on Leptogenesis}},  in \emph{{Proceedings
  of Theoretical Advanced Study Institute in Elementary Particle Physics :
  Exploring New Frontiers Using Colliders and Neutrinos (TASI 2006): Boulder,
  Colorado, June 4-30, 2006}}, pp.~123--176, 2007,
  \href{https://arxiv.org/abs/hep-ph/0703087}{{\ttfamily hep-ph/0703087}}.

\bibitem{Sakharov}
A.~D. Sakharov, \emph{{Violation of CP Invariance, C asymmetry, and baryon
  asymmetry of the universe}},
  \href{https://doi.org/10.1070/PU1991v034n05ABEH002497}{\emph{Pisma Zh. Eksp.
  Teor. Fiz.} {\bfseries 5} (1967) 32}.

\bibitem{Bell:2017irk}
N.~F. Bell, Y.~Cai, J.~B. Dent, R.~K. Leane and T.~J. Weiler, \emph{{Enhancing
  Dark Matter Annihilation Rates with Dark Bremsstrahlung}},
  \href{https://doi.org/10.1103/PhysRevD.96.023011}{\emph{Phys. Rev. D}
  {\bfseries 96} (2017) 023011}
  [\href{https://arxiv.org/abs/1705.01105}{{\ttfamily 1705.01105}}].

\bibitem{wellsannihilation}
J.~D. Wells, \emph{{Annihilation cross-sections for relic densities in the low
  velocity limit}},  \href{https://arxiv.org/abs/hep-ph/9404219}{{\ttfamily
  hep-ph/9404219}}.

\bibitem{kolbturner}
E.~W. Kolb and M.~S. Turner, \emph{{The Early Universe}}, {\emph{Front. Phys.}
  {\bfseries 69} (1990) 1}.

\bibitem{cpviolation}
M.~Garny, A.~Hohenegger and A.~Kartavtsev, \emph{Medium corrections to the
  $cp$-violating parameter in leptogenesis},
  \href{https://doi.org/10.1103/PhysRevD.81.085028}{\emph{Phys. Rev. D}
  {\bfseries 81} (2010) 085028}.

\bibitem{Mathematica}
{W}olfram Research{,}~Inc., \emph{Mathematica, {V}ersion 12.0},  Champaign, IL,
  (2019).

\bibitem{admeffectiveops}
M.~R. Buckley, \emph{Asymmetric dark matter and effective operators},
  \href{https://doi.org/10.1103/PhysRevD.84.043510}{\emph{Phys. Rev. D}
  {\bfseries 84} (2011) 043510}.

\bibitem{Sirunyan2019}
{\scshape CMS} collaboration, \emph{{Search for supersymmetry in proton-proton
  collisions at 13 TeV in final states with jets and missing transverse
  momentum}}, \href{https://doi.org/10.1007/JHEP10(2019)244}{\emph{Journal of
  High Energy Physics} {\bfseries 2019} (2019) }.

\bibitem{Aad:2019vvf}
{\scshape ATLAS} collaboration, \emph{{Search for direct production of
  electroweakinos in final states with one lepton, missing transverse momentum
  and a Higgs boson decaying into two $b$-jets in (pp) collisions at
  $\sqrt{s}=13$ TeV with the ATLAS detector}},
  \href{https://arxiv.org/abs/1909.09226}{{\ttfamily 1909.09226}}.

\bibitem{Aad:2019vnb}
{\scshape ATLAS} collaboration, \emph{{Search for electroweak production of
  charginos and sleptons decaying into final states with two leptons and
  missing transverse momentum in $\sqrt{s}=13$ TeV $pp$ collisions using the
  ATLAS detector}},  \href{https://arxiv.org/abs/1908.08215}{{\ttfamily
  1908.08215}}.

\bibitem{Benedikt:2018csr}
{\scshape FCC} collaboration, \emph{{FCC-hh: The Hadron Collider}},
  \href{https://doi.org/10.1140/epjst/e2019-900087-0}{\emph{Eur. Phys. J. ST}
  {\bfseries 228} (2019) 755}.

\bibitem{Krasnikov:1997nh}
N.~Krasnikov, \emph{{Influence of SU(2) x U(1) singlet scalars on Higgs boson
  signal at LHC}}, \href{https://doi.org/10.1142/S0217732398000978}{\emph{Mod.
  Phys. Lett. A} {\bfseries 13} (1998) 893}
  [\href{https://arxiv.org/abs/hep-ph/9709467}{{\ttfamily hep-ph/9709467}}].

\bibitem{Bowen:2007ia}
M.~Bowen, Y.~Cui and J.~D. Wells, \emph{{Narrow trans-TeV Higgs bosons and H
  ---> hh decays: Two LHC search paths for a hidden sector Higgs boson}},
  \href{https://doi.org/10.1088/1126-6708/2007/03/036}{\emph{JHEP} {\bfseries
  03} (2007) 036} [\href{https://arxiv.org/abs/hep-ph/0701035}{{\ttfamily
  hep-ph/0701035}}].

\bibitem{PhysRevD.89.083513}
L.~A. Anchordoqui, P.~B. Denton, H.~Goldberg, T.~C. Paul, L.~H.~M. da~Silva,
  B.~J. Vlcek et~al., \emph{Weinberg's higgs portal confronting recent lux and
  lhc results together with upper limits on ${B}^{+}$ and ${K}^{+}$ decay into
  invisibles}, \href{https://doi.org/10.1103/PhysRevD.89.083513}{\emph{Phys.
  Rev. D} {\bfseries 89} (2014) 083513}.

\bibitem{Curtin:2013fra}
D.~Curtin et~al., \emph{{Exotic decays of the 125 GeV Higgs boson}},
  \href{https://doi.org/10.1103/PhysRevD.90.075004}{\emph{Phys. Rev. D}
  {\bfseries 90} (2014) 075004}
  [\href{https://arxiv.org/abs/1312.4992}{{\ttfamily 1312.4992}}].

\bibitem{micromegasDD}
G.~Belanger, F.~Boudjema, A.~Pukhov and A.~Semenov, \emph{{Dark matter direct
  detection rate in a generic model with micrOMEGAs 2.2}},
  \href{https://doi.org/10.1016/j.cpc.2008.11.019}{\emph{Comput. Phys. Commun.}
  {\bfseries 180} (2009) 747}
  [\href{https://arxiv.org/abs/0803.2360}{{\ttfamily 0803.2360}}].

\bibitem{Bondarenko:2019vrb}
K.~Bondarenko, A.~Boyarsky, T.~Bringmann, M.~Hufnagel, K.~Schmidt-Hoberg and
  A.~Sokolenko, \emph{{Direct detection and complementary constraints for
  sub-GeV dark matter}},  \href{https://arxiv.org/abs/1909.08632}{{\ttfamily
  1909.08632}}.

\bibitem{Aalseth:2017fik}
C.~E. Aalseth et~al., \emph{{DarkSide-20k: A 20 tonne two-phase LAr TPC for
  direct dark matter detection at LNGS}},
  \href{https://doi.org/10.1140/epjp/i2018-11973-4}{\emph{Eur. Phys. J. Plus}
  {\bfseries 133} (2018) 131}
  [\href{https://arxiv.org/abs/1707.08145}{{\ttfamily 1707.08145}}].

\bibitem{Essig:2017kqs}
R.~Essig, T.~Volansky and T.-T. Yu, \emph{{New Constraints and Prospects for
  sub-GeV Dark Matter Scattering off Electrons in Xenon}},
  \href{https://doi.org/10.1103/PhysRevD.96.043017}{\emph{Phys. Rev.}
  {\bfseries D96} (2017) 043017}
  [\href{https://arxiv.org/abs/1703.00910}{{\ttfamily 1703.00910}}].

\bibitem{Akerib:2018dfk}
{\scshape LUX-ZEPLIN} collaboration, \emph{{Projected WIMP Sensitivity of the
  LUX-ZEPLIN (LZ) Dark Matter Experiment}},
  \href{https://arxiv.org/abs/1802.06039}{{\ttfamily 1802.06039}}.

\bibitem{superk}
{\scshape Super-Kamiokande Collaboration} collaboration, \emph{Search for
  proton decay via $p\ensuremath{\rightarrow}\ensuremath{\nu}{K}^{+}$ using
  $260\text{ }\text{
  }\mathrm{kiloton}\ifmmode\cdot\else\textperiodcentered\fi{}\mathrm{year}$
  data of super-kamiokande},
  \href{https://doi.org/10.1103/PhysRevD.90.072005}{\emph{Phys. Rev. D}
  {\bfseries 90} (2014) 072005}.

\bibitem{Migenda:2017oas}
{\scshape Hyper-Kamiokande Proto} collaboration, \emph{{The Hyper-Kamiokande
  Experiment: Overview \& Status}},  in \emph{{Proceedings, Prospects in
  Neutrino Physics (NuPhys2016): London, UK, December 12-14, 2016}}, 2017,
  \href{https://arxiv.org/abs/1704.05933}{{\ttfamily 1704.05933}}.

\bibitem{Acciarri:2016crz}
{\scshape DUNE} collaboration, \emph{{Long-Baseline Neutrino Facility (LBNF)
  and Deep Underground Neutrino Experiment (DUNE)}},
  \href{https://arxiv.org/abs/1601.05471}{{\ttfamily 1601.05471}}.

\bibitem{Abel:2020gbr}
{\scshape nEDM} collaboration, \emph{{Measurement of the permanent electric
  dipole moment of the neutron}},
  \href{https://doi.org/10.1103/PhysRevLett.124.081803}{\emph{Phys. Rev. Lett.}
  {\bfseries 124} (2020) 081803}
  [\href{https://arxiv.org/abs/2001.11966}{{\ttfamily 2001.11966}}].

\bibitem{Andreev:2018ayy}
{\scshape ACME} collaboration, \emph{{Improved limit on the electric dipole
  moment of the electron}},
  \href{https://doi.org/10.1038/s41586-018-0599-8}{\emph{Nature} {\bfseries
  562} (2018) 355}.

\bibitem{Binosi:2008ig}
D.~Binosi, J.~Collins, C.~Kaufhold and L.~Theussl, \emph{{JaxoDraw: A Graphical
  user interface for drawing Feynman diagrams. Version 2.0 release notes}},
  \href{https://doi.org/10.1016/j.cpc.2009.02.020}{\emph{Comput. Phys. Commun.}
  {\bfseries 180} (2009) 1709}
  [\href{https://arxiv.org/abs/0811.4113}{{\ttfamily 0811.4113}}].

\bibitem{Riotto:1998bt}
A.~Riotto, \emph{{Theories of baryogenesis}},  in \emph{{Proceedings, Summer
  School in High-energy physics and cosmology: Trieste, Italy, June 29-July 17,
  1998}}, pp.~326--436, 1998,
  \href{https://arxiv.org/abs/hep-ph/9807454}{{\ttfamily hep-ph/9807454}}.

\bibitem{Buchmuller:2005eh}
W.~Buchmuller, R.~D. Peccei and T.~Yanagida, \emph{{Leptogenesis as the origin
  of matter}},
  \href{https://doi.org/10.1146/annurev.nucl.55.090704.151558}{\emph{Ann. Rev.
  Nucl. Part. Sci.} {\bfseries 55} (2005) 311}
  [\href{https://arxiv.org/abs/hep-ph/0502169}{{\ttfamily hep-ph/0502169}}].

\end{thebibliography}\endgroup

\end{document}